%% file: shell.tex
\documentclass{aa}
\usepackage{times}
\usepackage{array}
\usepackage{epsfig}
\usepackage{amssymb}
\usepackage{amsmath}
\usepackage{longtable}
\usepackage{subfigure}
\input ndef.tex

\begin{document}
\newcommand{\Msun}{\ensuremath{\mathrm{M}_\odot}}
\newcommand{\Lsun}{\ensuremath{\mathrm{L}_\odot}}
\newcommand{\Rsun}{\ensuremath{\mathrm{R}_\odot}}
\newcommand{\Mdot}{\ensuremath{\dot{M}}}
\newcommand{\arrow}{$\longrightarrow$}
\newcommand{\lsim}{\mathrel{\hbox{\rlap{\lower.55ex \hbox {$\sim$}}
 \kern-.3em \raise.4ex \hbox{$<$}}}}
\newcommand{\gsim}{\mathrel{\hbox{\rlap{\lower.55ex \hbox {$\sim$}}
 \kern-.3em \raise.4ex \hbox{$>$}}}}
\newcommand{\half}{\ensuremath{{\textstyle\frac{1}{2}}}}
\newcommand{\e}{\ensuremath{\mathrm{e}}}
\newcommand{\n}{\ensuremath{\mathrm{n}}}
\newcommand{\p}{\ensuremath{\mathrm{p}}}
\renewcommand{\H}[1]{\ensuremath{{^{#1}\mathrm{H}}}}
\newcommand{\D}{\ensuremath{\mathrm{D}}}
\newcommand{\He}[1]{\ensuremath{{^{#1}\mathrm{He}}}}
\newcommand{\Li}[1]{\ensuremath{{^{#1}\mathrm{Li}}}}
\newcommand{\Be}[1]{\ensuremath{{^{#1}\mathrm{Be}}}}
\newcommand{\B}[1]{\ensuremath{{^{#1}\mathrm{B}}}}
\newcommand{\C}[1]{\ensuremath{{^{#1}\mathrm{C}}}}
\newcommand{\N}[1]{\ensuremath{{^{#1}\mathrm{N}}}}
\renewcommand{\O}[1]{\ensuremath{{^{#1}\mathrm{O}}}}
\newcommand{\F}[1]{\ensuremath{{^{#1}\mathrm{F}}}}
\newcommand{\Na}[1]{\ensuremath{{^{#1}\mathrm{Na}}}}
\newcommand{\Ne}[1]{\ensuremath{{^{#1}\mathrm{Ne}}}}
\newcommand{\Mg}[1]{\ensuremath{{^{#1}\mathrm{Mg}}}}
\newcommand{\Fe}[1]{\ensuremath{{^{#1}\mathrm{Fe}}}}
\newcommand{\Ni}[1]{\ensuremath{{^{#1}\mathrm{Ni}}}}
\newcommand{\Co}[1]{\ensuremath{{^{#1}\mathrm{Co}}}}
\newcommand{\Al}[1]{\ensuremath{{^{#1}\mathrm{Al}}}}
\newcommand{\avg}[1]{\ensuremath{\langle#1\rangle}}
\newcommand{\rate}[1]{\ensuremath{r_\mathrm{#1}}}
\newcommand{\sv}[1]{\ensuremath{\langle\sigma v\rangle_\mathrm{#1}}}
\newcommand{\life}[2]{\ensuremath{\tau_\mathrm{#1}(\mathrm{#2})}}
\newcommand{\fpp}[1]{\ensuremath{f_\mathrm{pp#1}}}
\newcommand{\TW}{\ensuremath{E_{\rm rot}/|E_{\rm grav}|}}

\newcommand{\hemdota}{$2\times10^{-7}$}
\newcommand{\hemdotb}{$3\times10^{-7}$}
\newcommand{\hemdotc}{$5\times10^{-7}$}
\newcommand{\hemdotd}{$10^{-6}$}
\newcommand{\Mwd}{$M_{\rm WD}$}
\newcommand{\Minit}{$M_{\rm init}$}
\newcommand{\msyr}{$\rm{M}_{\odot}/\rm{yr}$}
\newcommand{\kmpersec}{$\rm{km}~\rm{s}^{-1}$}
\newcommand{\ctw}{$^{12}\rm{C}$}
\newcommand{\ost}{$^{16}\rm{O}$}
\newcommand{\rtrip}{$3\alpha$}
\newcommand{\rche}{$^{12}\rm{C}(\alpha,\gamma)^{16}\rm{O}$}

\authorrunning{Yoon, Langer \& Scheithauer}
\titlerunning{Effects of rotation on the helium shell source in WDs} 

\title{Effects of rotation on the helium burning shell source in accreting white dwarfs}

\author{S.-C. Yoon\inst{1}, N. Langer\inst{1} \and S. Scheithauer\inst{2}}

\institute{
Astronomical Institute, Utrecht University, Princetonplein 5,
NL-3584 CC, Utrecht, The Netherlands
\and
Institut f\"ur Physik, Universit\"at
Potsdam, D--14415~Potsdam, Germany
}

\offprints {\tt S.C.Yoon@astro.uu.nl}
\date{Received  ; accepted , }

\abstract{
We investigate the effects of rotation on the behavior of the helium burning shell source 
in accreting carbon-oxygen white dwarfs, 
in the context of the single degenerate Chandrasekhar mass progenitor scenario 
for Type~Ia supernovae (SNe~Ia). We model the evolution of helium accreting white
dwarfs of initially 1 \Msun{}, assuming four different constant accretion rates 
(2, 3, 5 and 10$\times10^{-7}$ \msyr). In a one-dimensional
approximation, we compute the mass accretion and subsequent nuclear
fusion of helium into carbon and oxygen, as well as angular momentum accretion,
angular momentum transport inside the white dwarf, and rotationally induced
chemical mixing. Our models show two major effects of rotation:
{\bf a)} The helium burning nuclear shell source in the rotating models
is much more stable than in corresponding non-rotating models ---
which increases the likelihood of accreting white dwarfs to reach the stage
of central carbon ignition. 
This effect is mainly due to rotationally induced mixing at the CO/He
interface which widens the shell source, and due to the centrifugal force
lowering the density and degeneracy at the shell source location.
{\bf b)} The C/O-ratio in the layers which experience helium shell burning
--- which may affect the energy of a  SN~Ia explosion --- is
strongly decreased by the rotationally induced mixing of 
$\alpha$-particles into the carbon-rich layers.
We discuss implications of our results for the evolution of SNe Ia progenitors.
}

\maketitle

\section{Introduction}

Accreting white dwarfs in close binary systems are considered to be responsible 
for such important astrophysical phenomena as nova explosions, 
cataclysmic variables and super-soft X-ray sources.  
Especially, there is general agreement that accreting 
CO white dwarfs are progenitors of Type~Ia supernovae (SNe~Ia)
(e.g., Woosley \& Weaver~\cite{Woosley86}; Branch et al.~\cite{Branch95}; 
Wheeler~\cite{Wheeler96}; Nomoto et al.~\cite{Nomoto97b}, 
Hillebrandt \&{} Niemeyer~\cite{Hillebrandt00}; Livio~\cite{Livio01}).

The fairly homogeneous properties of light curves and spectra of SNe~Ia have made 
them a valuable distance indicator to determine cosmological 
parameters (e.g, Hamuy et al.~\cite{Hamuy96a}; Branch~\cite{Branch98}; Leibundgut~\cite{Leibundgut01}).  
One of the most intriguing results obtained from SNe~Ia data may be the recent suggestion of a
finite cosmological constant, which implies that the universe is currently accelerating its 
expansion (Perlmutter et al.~\cite{Perlmutter99a}; Riess et al.~\cite{Riess00}).  

The observational properties of SNe~Ia demand a thermonuclear explosion of
a CO white dwarf as their origin. 
However, despite the far-reaching observational achievements, 
self-consistent progenitor evolution model calculations which explain the existence
of SNe~Ia are still lacking (but see Yoon \& Langer~\cite{Yoon03}). 
Models for the evolution of SN~Ia progenitors studied until now can be 
divided into two groups, one where the white dwarf explosion is
triggered close to the center, and the other where the white dwarf
explosion is triggered in the helium shell
(see Livio~\cite{Livio01} for a review). 

Two scenarios exist for close-to-center ignition. 
One is the double degenerate scenario, in which two CO white dwarfs in a binary system
merge due to the angular momentum loss by gravitational wave radiation. 
If the system mass exceeds the Chandrasekhar mass, carbon must ignite,
with the possible outcome of a thermonuclear explosion.
Although there is observational evidence for double white dwarf systems 
(e.g., Koester et al.~\cite{Koester01}; Karl et al.~\cite{Karl03}), 
including one sdB star plus a CO white dwarf system which might
result in a double white dwarf system having a total mass beyond the Chandrasekhar limit 
(Maxted et al.~\cite{Maxted00}), 
this scenario is criticized mainly by the fact
that the merging of two white dwarfs is likely to end in an accretion-induced
collapse rather than a  SN explosion (Saio \& Nomoto~\cite{Saio85},~\cite{Saio98}).

The currently favored scenario for close-to-center ignition is the
single degenerate scenario, in which a CO white dwarf in a close binary system
accretes hydrogen or helium rich matter from a non-degenerate companion at
a relatively high accretion rate ($\sim$ $10^{-7}$\msyr{}).
The super-soft X-ray sources, which are believed to consist of a non-degenerate
star and a white dwarf, are promising observed counterparts
of this scenario. The idea of steady nuclear burning due to a 
high accretion rate fits well to these sources (e.g., Li \& van den Heuvel~\cite{Li97}; 
Kahabka \& van den Heuvel~\cite{Kahabka97}; Langer et al.~\cite{Langer00}; 
Yoon \& Langer~\cite{Yoon03}).
Symbiotic systems (e.g.,  Hachisu et al.~\cite{Hachisu99}) and recurrent novae
(e.g.,  Hachisu \& Kato~\cite{Hachisu01}; Thoroughgood et al.~\cite{Thoroughgood01}) 
may also belong to this scenario. 
However, it is still not clear whether this scenario
can explain the observed SN~Ia frequency since the accretion rates which 
allow steady burning are limited to a very narrow range.
Furthermore, helium shell burning is generally found to be unstable: 
strong helium shell flashes might lead to strong mass loss by
optically thick winds, or to a strong expansion of the
white dwarf envelope to a red-giant phase which will result in a possible merging of the binary 
(e.g. Nomoto~\cite{Nomoto82a}; Hachisu, Kato \& Nomoto~\cite{Hachisu96}; Cassisi et al.~\cite{Cassisi98}; 
Kato \& Hachisu~\cite{Kato99}; Langer et al.~\cite{Langer02}). 

In the  so called `Edge Lit Detonation' models, corresponding to
sub-Chandrasekhar mass models, helium ignites at the bottom of 
a helium layer of a few tenth of a solar mass, accumulated on a CO white dwarf
at a  relatively low accretion rate ($\sim 10^{-8}{\rm M_{\odot}/yr}$)
(e.g. Nomoto~\cite{Nomoto82b}; Limongi \& Tornamb\'e~\cite{Limongi91};  
Woosley \& Weaver~\cite{Woosley94}; Livne \& Arnett~\cite{Livne95}). 
These models are favored in terms of
statistics (e.g. Iben \& Tutukov~\cite{Iben91}; Tout et al.~\cite{Tout01}) but disfavored by the fact that 
the spectra and light curves produced by these models are
not in agreement with observations (H\"oflich et al.~\cite{Hoeflich96b}; Nugent et al.~\cite{Nugent97}; 
Pinto et al.~\cite{Pinto01}). Yoon \& Langer (\cite{Yoon04b}) 
showed that
a helium detonation in such models may be avoided altogether due to a strong shear layer 
inside an accreting white dwarf. 

Most previous studies of accreting white dwarfs as SN~Ia progenitors have 
assumed spherical symmetry. 
However, in a close binary system which contains a white dwarf,  
an accretion disk is formed around the white dwarf.
The matter coming from the accretion disk 
which is incorporated into the white dwarf  
carries high specific angular momentum, i.e. up to the value
corresponding to Keplerian rotation at the white dwarf equator. 
Consequently, not only the accreted layer may be spinning very rapidly,
but also the rest of the white dwarf, as
angular momentum may be transported
into the stellar interior (e.g. Ritter~\cite{Ritter85};  Langer et al.~\cite{Langer00}).

Recent polarization observations provide evidence 
for the aspherical nature of SNe~Ia
(SN 1999by, Howell et al.~\cite{Howell01}; SN 2001el, Wang et al~\cite{Wang03}). 
A plausible explanation for this asphericity may be a rapidly rotating 
progenitor, while clumpiness of supernova ejecta or binarity of the
progenitors could be
another possible cause for the observed polarization (Kasen et al.~\cite{Kasen03}).
The observation that white dwarfs in Cataclysmic Variables rotate much faster than
isolated ones (Sion~\cite{Sion99}; Starrfield~\cite{Starrfield03}) supports the idea
that accreting white dwarfs are indeed spun up. 
In this picture, differential rotation inside the
accreting white dwarf is inevitable, unless the angular momentum 
transport in the white dwarf is extremely efficient such that
rigid body rotation may be maintained throughout the white dwarf interior
(Yoon \& Langer~\cite{Yoon04a}). 
Rotationally induced instabilities such as the shear instability
may lead to chemical mixing 
between the CO white dwarf material and the accreted matter. 
This effect has been discussed
by several authors mainly with respect to the chemical
enrichment of CNO and other heavier elements observed in nova
ejecta (e.g., Kippenhahn \&{} Thomas~\cite{Kippenhahn78}; MacDonald~\cite{MacDonald83};
Livio \&{} Truran~\cite{Livio87}; Fujimoto~\cite{Fujimoto88}; Fujimoto \&{} Iben~\cite{Fujimoto97}). 

In this paper, we investigate
the effects of rotation on the evolution of accreting white dwarfs 
in relation to the single degenerate Chandrasekhar mass scenario for SNe Ia progenitors. 
For this purpose, we carry out numerical simulations of  helium burning white dwarfs 
considering high accretion rates ($>10^{-7}$\msyr) which allow steady shell burning
for hydrogen or helium. 
Our discussions in this study are focused on the rotationally induced
chemical mixing and their consequences for the behavior of the helium burning shell source.
Comprehensive discussions on the rotational induced hydrodynamic instabilities
and their role in the redistribution of angular momentum in accreting
white dwarfs are given in a separate paper (Yoon \& Langer~\cite{Yoon04a}).
Effects of rotation on  sub-Chandrasekhar mass models, i.e., on accreting 
white dwarfs with relatively low accretion rates ($\sim10^{-8}$ \msyr), 
are discussed in Yoon \& Langer (\cite{Yoon04b}).   
Our computational method is introduced in Sect.~2, and
the initial model and physical assumptions are discussed in Sect.~3.
In Sect.~4, we present the evolution of various physical quantities of 
our white dwarf models. We discuss rotationally 
induced chemical mixing and the stability of shell burning in 
Sect.~5 and~6 respectively.  
Finally, we discuss our results in Sect.~7.

\section{Method}\label{sect:method}

The numerical models have been computed with a hydrodynamic stellar evolution
code (cf, Langer~\cite{Langer98}; Heger, Langer \&{} Woosley~\cite{Heger00a}). 
Opacities are taken from Iglesias \& Rogers (\cite{Iglesias96}). 
Changes in the chemical composition
are computed using a nuclear network with more than 60 nuclear
reactions, of which 31 reactions are
relevant to helium burning such as \rtrip{}, 
$^{12}\rm{C}(\alpha,\gamma)^{16}\rm{O}(\alpha,\gamma)^{20}\rm{Ne}$ and
$^{20}\rm{Ne}(\alpha,\gamma)^{24}Mg(\alpha,\gamma)^{28}\rm{Si}$ reactions. 
For the \rche{} reaction rate, which has an important role in our study, 
we use the rate of Caughlan et al. (\cite{Caughlan85}) reduced by a factor of 0.63, 
as suggested by Weaver \& Woosley (\cite{Weaver93}).

The effect of the centrifugal force on the stellar structure,
and rotationally induced transport of angular momentum and chemical species
are treated in a one-dimensional approximation (Heger et al.~\cite{Heger00a}).   
We consider rotationally induced transport processes as
Eddington-Sweet circulations, the Goldreich-Schubert-Fricke
instability, and the dynamical and secular shear instability, 
by solving a non-linear diffusion equation, with  diffusion coefficients
for each of the instabilities computed as in Heger et al. (\cite{Heger00a})
and Yoon \& Langer (\cite{Yoon04a}). 
The viscous dissipation of the rotational energy
is also considered as described by Yoon \& Langer (\cite{Yoon04a}).
More details about the
numerical methods are given in Yoon \& Langer (\cite{Yoon04a}).

A comparison of our method with previously used ones for rotating white dwarf models
is already  given in Yoon \& Langer (\cite{Yoon04a}) in detail. 
Here we give some additional comments on the rotation physics 
in relation to helium accreting white dwarfs.

Our approach to include rotation is based on the method given 
by Kippenhahn \&{} Thomas (\cite{Kippenhahn70}),
with the mass shells of a star defined by isobars instead of spherical shells. 
In describing effects of rotationally induced chemical mixing, 
the chemical abundances are  assumed to be constant on isobars.
This is justified by theoretical achievements of Zahn (\cite{Zahn75}), Chaboyer \& Zahn (\cite{Chaboyer92})
and Zahn (\cite{Zahn92}), who showed that anisotropic turbulence acts much more efficiently 
in the horizontal than in the vertical direction. Shellular rotation 
and chemical homogeneity  on isobars  can be thus achieved (Meynet \& Maeder~\cite{Meynet97}), enabling
us to use a one dimensional approximation (see also Maeder~\cite{Maeder03a}). 
 
Our assumption of shellular rotation may be questionable for 
helium accreting white dwarfs, on two accounts.
Kippenhahn \& Thomas (\cite{Kippenhahn78}) argued that the accreted matter
cannot spread over the whole surface of white dwarf but rather forms an accretion belt,
since the angular momentum transport time scale based on the electron viscosity
is longer than the accretion time scale.
However, calculations by Sparks \&{} Kutter (\cite{Sparks87})
failed in obtaining classical nova outbursts when the prescription of accretion belts
was employed. Livio \&{} Truran (\cite{Livio87}) argued that if a turbulent 
viscosity is used, the angular momentum transport becomes much more efficient
than discussed by Kippenhahn \&{} Thomas. 
They concluded that the accreted matter 
can spread over the entire star with the consequence that
mass accretion onto white dwarfs may result in nova outbursts, 
which was not possible in the accretion belt prescription due to
the too strong centrifugal force at the equator. 
A local stability analysis by MacDonald (\cite{MacDonald83}) and a theoretical study 
by Fujimoto (\cite{Fujimoto88}) also concluded
that the accreted matter will not remain confined to the equatorial belt
due to efficient angular momentum transport toward the poles.
We therefore conclude that the assumption of shellular rotation 
employed in our study may still be well-grounded, 
at least in the non-degenerate helium envelope,
although multi-dimensional simulations are ultimately required
for its justification (cf. Hujeirat 1995). 

Secondly, 
as degenerate matter is barotropic, perturbations which create
$\partial j/\partial z \ne 0$, with $j$ as specific angular momentum
and $z$ being the distance to 
the equatorial plane, are eliminated on a short time scale. Thus,
cylindrical rotation may be enforced in the inner CO core 
(e.g. Kippenhahn \& M\"ollenhoff~\cite{Kippenhahn74}; Durisen~\cite{Durisen77}). 
This means that accreting white dwarfs may consist of two different
rotation laws: shellular in the non-degenerate envelope and 
cylindrical rotation in the degenerate core. 
However, our numerical models can still represent 
the cylindrically rotating degenerate inner core to some degree, 
since most of the total angular momentum is confined
at the equatorial plane in both the shellular and the cylindrical cases.
It is harder to estimate the angular momentum transport efficiency 
in the semidegenerate transition layer, as non-barotropic perturbations in the degenerate core may
induce dynamical meridional flows 
(Kippenhahn \& M\"ollenhoff~\cite{Kippenhahn74}; M\"uller~2003, private communication).
Since it is poorly known 
how these meridional flows may affect the angular momentum redistribution,
we have to leave the detailed investigation of this effect to future work. 

Finally, the accuracy of our results for very rapid
rotation is limited due to the one dimensional approximation.
In computing the effective gravitational potential in a rotating star,
the potential is expanded in terms of spherical harmonics,
of which we only consider terms up to the second order
(Kippenhahn \& Thomas~\cite{Kippenhahn70}; Endal \& Sofia~\cite{Endal76}).
This method can
reproduce the shapes of rigidly rotating polytropes accurately
up to a rotation rate of about 60\% of the critical value, corresponding
to correction factors of $f_P\simeq 0.75$ and $f_T \simeq 0.95$
in the stellar structure equations (Fliegner~\cite{Fliegner93}; Heger et al.~\cite{Heger00a}).
We therefore limit these factors to the quoted values,
with the consequence that we underestimate the effect
of the centrifugal force in layers which rotate more rapidly than
about 60\% critical. We shall see below that this affects
only a very small fraction of our stellar models in most cases. 
However,
as the surface layers are always very close to critical rotation,
the stellar radius of our models may be somewhat underestimated.
A better description of the equilibrium structure of rapidly rotating stars which
may deviate significantly from spheres
requires multi-dimensional techniques
such as the self-consistent field method (Ostriker \& Mark~\cite{Ostriker68b}; Hachisu~\cite{Hachisu86}).
In addition, since our models are rotationally symmetric, 
three dimensional effects such as triaxial
deformation can not be described with our numerical code.
 
\section{Physical assumptions}

\begin{table*}
\begin{center}
\caption{Properties of the computed models. The first column denotes 
the sequence number, where 'NR', 'R' and 'RT' denote
the non-rotating models and rotating models with $f_{\rm c} =$ 1/30 and 1/3 respectively. 
The second column gives the accretion rate in the unit of $10^{-7}$ \msyr{}. 
The third column denotes the factor $f_{\rm c}$ for the chemical mixing efficiency in rotating models.
Columns~4, 5, 6 and~7 give mass, central density, central temperature and \TW{} of the last computed model 
of each evolutionary sequence. The eighth column gives the amount of accreted
material until the onset of the helium shell instability.
The column with $L_{\rm Edd}$ indicates whether the last computed model reached the Eddington limit or not.
The last column gives the total number of the computed models for each sequence.
}\label{tab1}
\begin{tabular}{c c c c c c c c c r}
\hline \hline
No.  & \Mdot{} & $f_{\rm c}$& $M_{\rm last}$ & $\rho_{\rm c, last}$ & $T_{\rm c, last}$ & $(E_{\rm rot}/|E_{\rm grav}|)_{\rm last}$ & $\Delta M$ &  $L_{\rm Edd}$ &  No. of Models   \\
     & [$10^{-7}$ \msyr]   &    &   [\Msun]         &     [$10^8~{\rm g/cm^3}$]  & [$10^8$ K] &   & [\Msun] &  \\
\hline
NR1  & 2.0 & -- & 1.0026 & 0.25 & 1.45 & -- & 0.00 & Yes & 4342  \\
NR2  & 3.0 & -- & 1.0042 & 0.25 & 1.45 & -- & 0.00 & Yes & 5164 \\
NR3  & 5.0 & -- & 1.0130 & 0.28 & 1.37 & -- & 0.00 & Yes & 10324 \\
NR4  &  10 & -- & 1.1984 & 1.39 & 1.44 & -- & 0.03 &  No & 383637  \\
\hline
R1   & 2.0 & 1/30 & 1.0074 & 0.28 & 1.25 & 0.0018 & 0.00 & Yes & 18021 \\ 
R2   & 3.0 & 1/30 & 1.0224 & 0.32 & 1.17 & 0.0025 & 0.00 & Yes & 28449 \\
R3   & 5.0 & 1/30 & 1.2241 & 1.05 & 1.21 & 0.0197 & 0.14 &  No & 180004\\
R4   &  10 & 1/30 & 1.4773 & 5.14 & 1.72 & 0.0578 & 0.48 &  No & 490026\\
RT1   & 2.0 & 1/3 & 1.0665 & 0.42 & 0.99 & 0.0081 & 0.01 &  No & 174270\\
RT2   & 3.0 & 1/3 & 1.1537 & 0.77 & 1.22 & 0.0080 & 0.01 &  No & 212730\\
\hline
\end{tabular}
\end{center}
\end{table*}

We begin with a hot carbon-oxygen white dwarf 
of $\log L_{\rm s}/\rm{L_{\odot}}= 3.379$ in order to avoid
the numerical difficulty of a strong initial helium shell flash.   
The initial mass of our model is 0.998 \Msun. 
The central temperature and density 
are initially such that $T_{\rm c}=2.288\times10^8~{\rm K}$ 
and $\rho_{\rm c}=1.64\times10^6~\rm{g~cm^{-3}}$, 
respectively. 
Although in reality white dwarfs may be much colder when mass accretion starts,
the thermal evolution of the white dwarf core is not influenced by the
different initial conditions due to the self heating of white dwarfs 
(cf. Yoon \& Langer~\cite{Yoon03}).
Since observations (Heber et al.~\cite{Heber97}; Koester et al.~\cite{Koester98}; Kawaler~\cite{Kawaler03})
and stellar evolution models (Langer et al.~\cite{Langer99}) find small spin rates of isolated white dwarfs,
we adopt a slow and rigid rotation in our initial model: the initial rotation velocity 
at the white dwarf equator is set to 1.0 \kmpersec. 

We consider 4 different accretion rates: 
\hemdota, \hemdotb, \hemdotc{} and \hemdotd{} \msyr.
The accreted matter is assumed to be helium-rich with $Y=0.98$ and $Z=0.02$.
The abundances for heavier elements are chosen to be the same as
in a helium core which results from core hydrogen burning in a 
main sequence star through the CNO cycle. 

The specific angular momentum of the accreted matter is assumed to have the
local Keplerian value, as most white dwarfs with a non-degenerate companion 
are believed to accrete matter through an accretion disk. 
However, analytic considerations can show this to cause an angular momentum problem. 
Assuming rigid rotation and a usual white dwarf mass-radius relation, 
the white dwarf reaches overcritical rotation
much before reaching the Chandrasekhar limit
(Livio \& Pringle~\cite{Livio98}; Langer et al.~\cite{Langer00}). 
For finite angular momentum redistribution times, as realized by
our preliminary calculations (Yoon \& Langer~\cite{Yoon02}), 
overcritical rotation is even reached earlier.

Livio \& Pringle (\cite{Livio98}) suggested nova explosions as a possible mechanism
for angular momentum loss to explain the white dwarf rotation velocity in typical CV systems.
However, within the single degenerate SNe~Ia progenitor scenario in which
the white dwarf must grow to the Chandrasekhar limit by mass accretion, 
we need another mechanism to solve the angular momentum
problem. To prevent the white dwarf from rotating over-critically,  
in our calculations, we limited the angular momentum gain 
in a simple way as follow. 
If the surface velocity at the white dwarf equator
is below the Keplerian value, the specific angular momentum of the accreted matter
is assumed to have the Keplerian value. Otherwise, no angular 
momentum is allowed to be accreted onto the white dwarf while mass is still accreted: 
\begin{equation}\label{eq:1}
j(\rm accreted~matter) = \left\{ \begin{array}{ll} 
v_{\rm{Kepler}} R_{\rm{WD}} & \textrm{if $ v_{\rm{s}} < v_{\rm{Kepler}}$ } \\
0 &  \textrm{if $ v_{\rm{s}} = v_{\rm{Kepler}}$ }~, \end{array} \right.
\end{equation}
where $v_{\rm{Kepler}} = \sqrt{GM_{\rm{WD}}/R_{\rm{WD}}}$, and $v_{\rm{s}}$
is the surface rotation velocity of the accreting white dwarf at its equator. 
This assumption may be justified by considerations of Paczy\'nski (\cite{Paczynski91}) 
and Popham \& Narayan (\cite{Popham91}) to allow angular momentum transport from the white dwarf
surface through the boundary layer into the accretion disk by turbulent viscosity 
when the spin rate of the white dwarf reaches the Keplerian value, without preventing efficient
mass accretion.

As discussed in Heger et al. (\cite{Heger00a}), the contribution of the rotationally
induced instabilities to the chemical diffusion is considered 
by introducing a factor $f_{\rm c}$ $\in$ [0, 1], which 
describes the ratio of  the chemical diffusion coefficient to turbulent viscosity.
Heger et al. (\cite{Heger00a}) and Heger \& Langer (\cite{Heger00b}) chose a value of $f_{\rm c} = 1/30$
following Chaboyer \& Zahn (\cite{Chaboyer92}) (See also Pinsonneault et al.~\cite{Pinsonneault89}). 
In the present study, we consider two different values for $f_{\rm c}$, i.e., 1/30 and
1/3 (see Table~\ref{tab1}), in order to investigate how the strength of chemical mixing affects
the nuclear shell burning.

For a comparative study, rotating and non-rotating
models with the same initial conditions have been investigated.  
The properties of all computed models are summarized in Table~\ref{tab1}.

\section{Evolution}\label{sect:evolution}

\begin{figure}[t]
\epsfxsize=\hsize
\epsffile{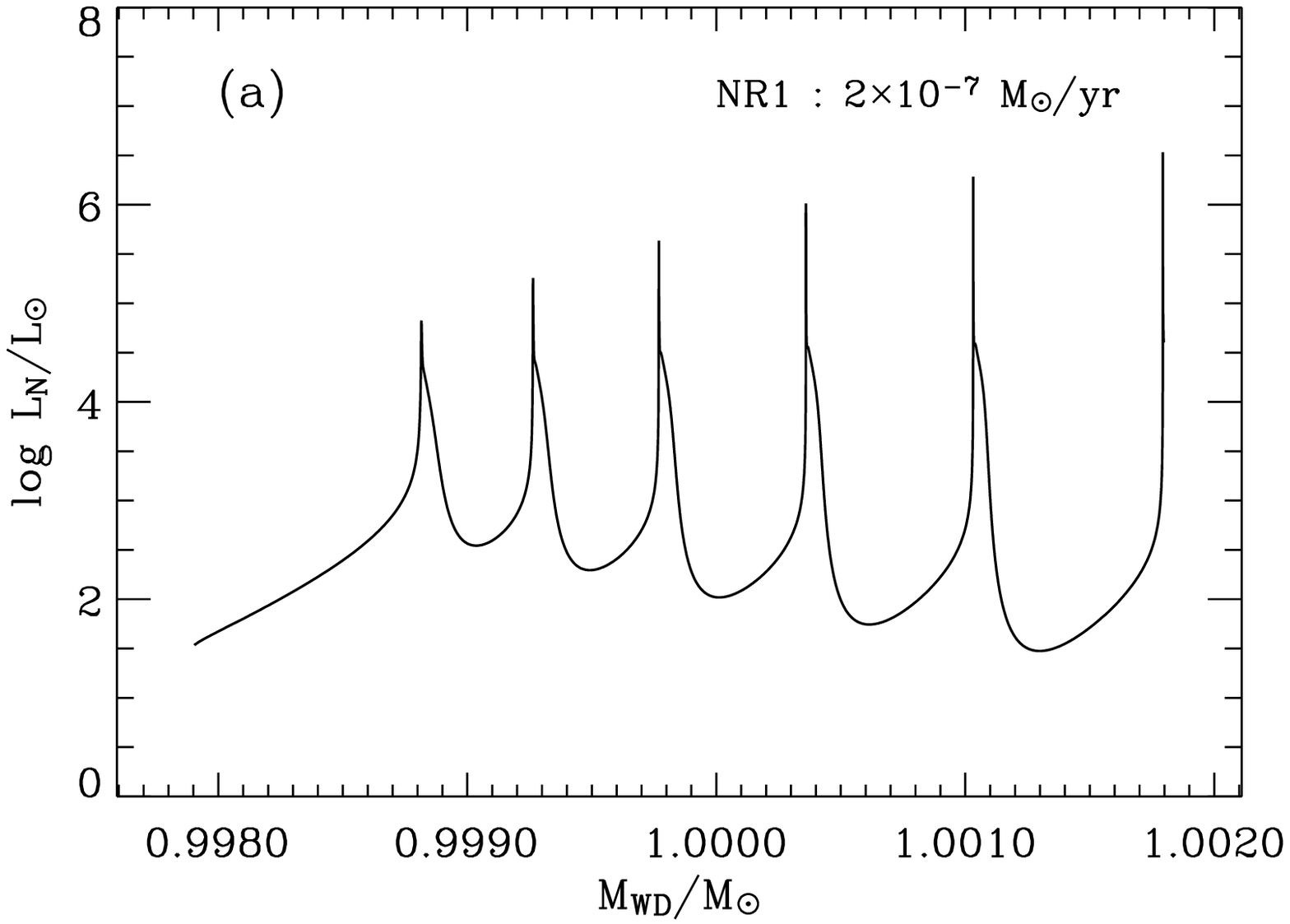}
\epsfxsize=\hsize
\epsffile{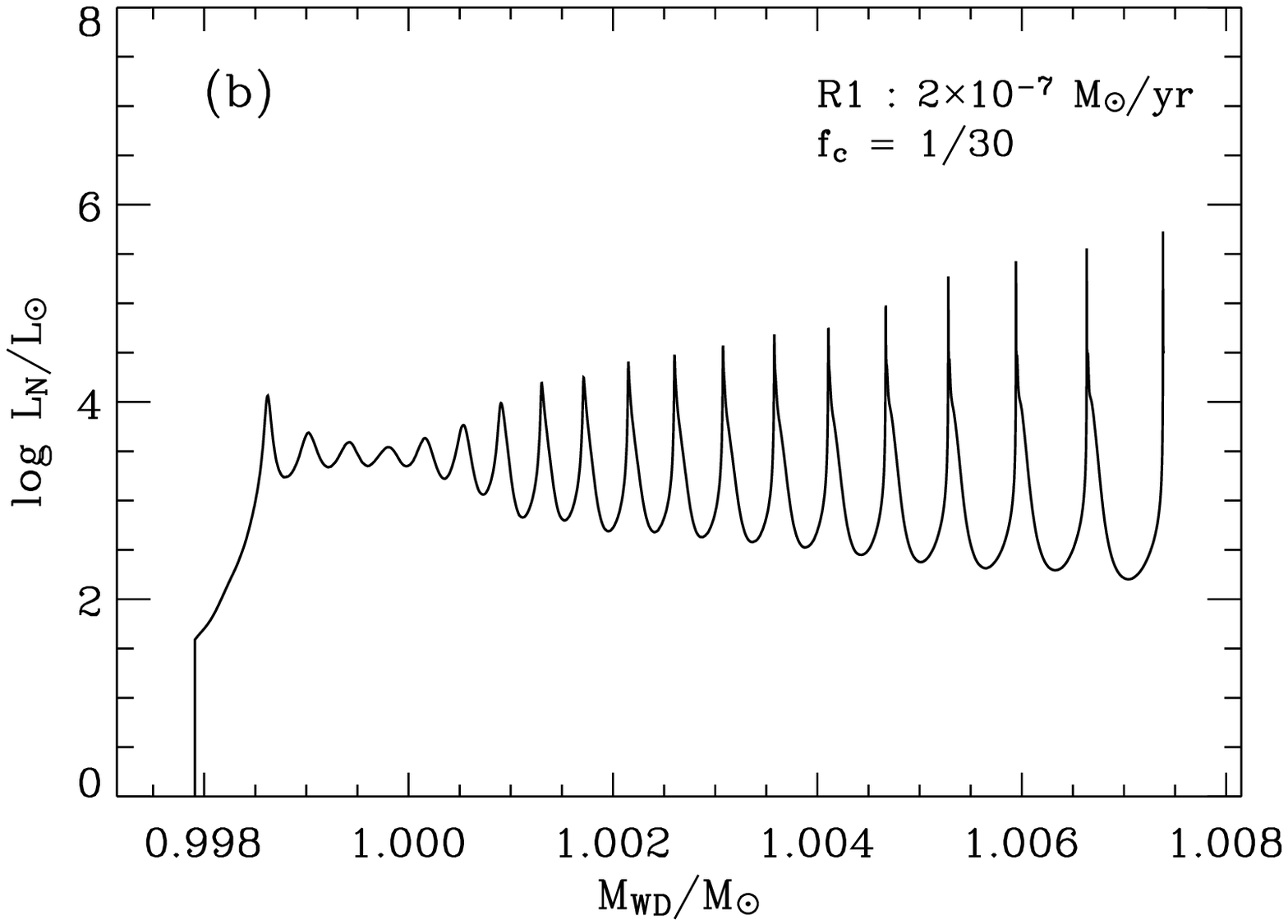}
\epsfxsize=\hsize
\epsffile{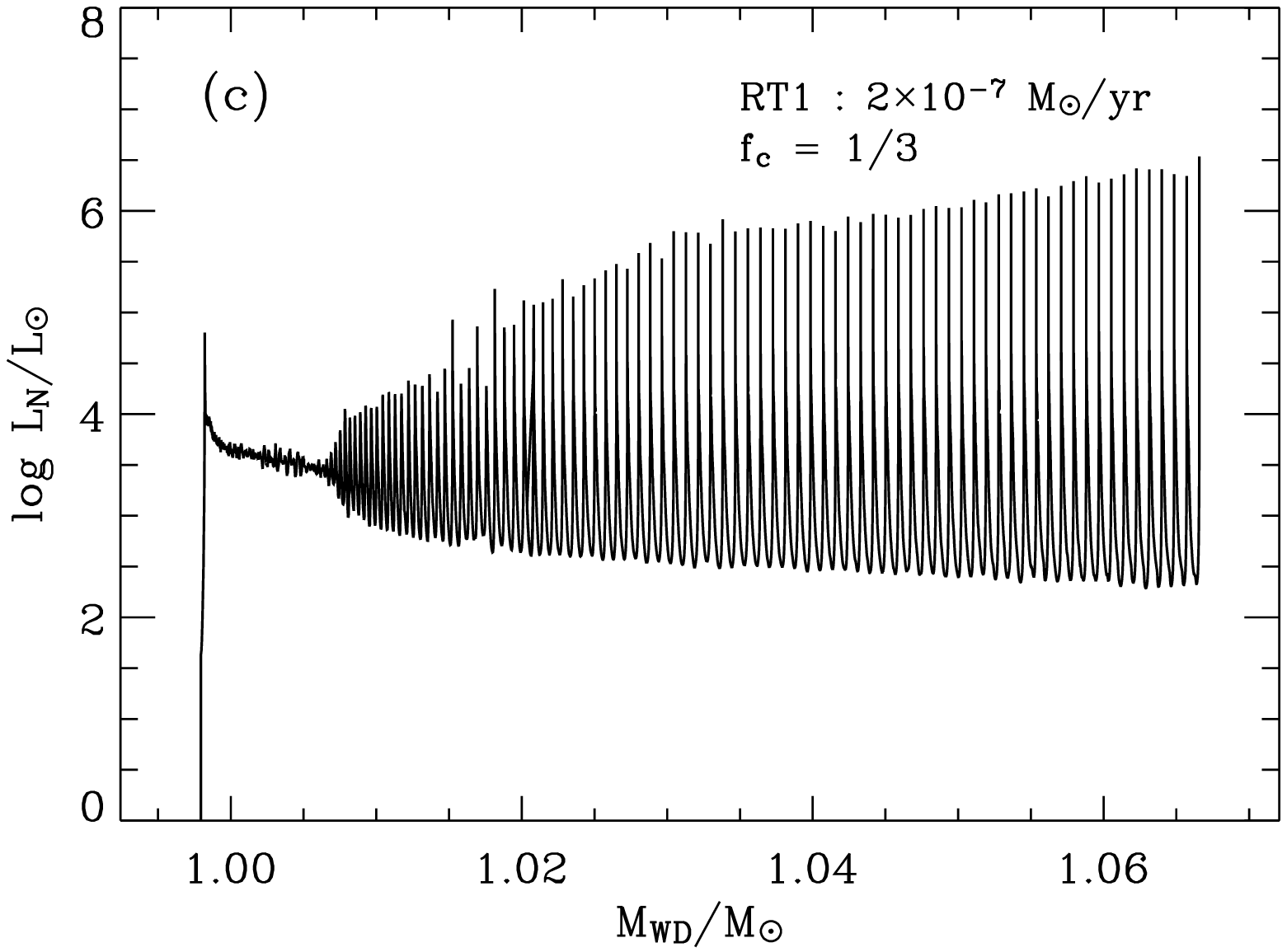}
\caption{Evolution of nuclear luminosity due to helium burning in the helium accreting white
dwarf models with \Mdot{} = \hemdota{}~\msyr. 
The box in panel (a)  gives 
the result from the non-rotating case (NR1)
while the results from the rotating models with $f_{\rm c}=1/30$ (R1) and $f_{\rm c}=1/3$ (RT1)
are shown in panels (b) and (c).
The abscissa denotes the total mass of white dwarf, which serves as a measure of time.
}\label{fig:lhe1}
\end{figure}

\begin{figure}[t]
\epsfxsize=\hsize
\epsffile{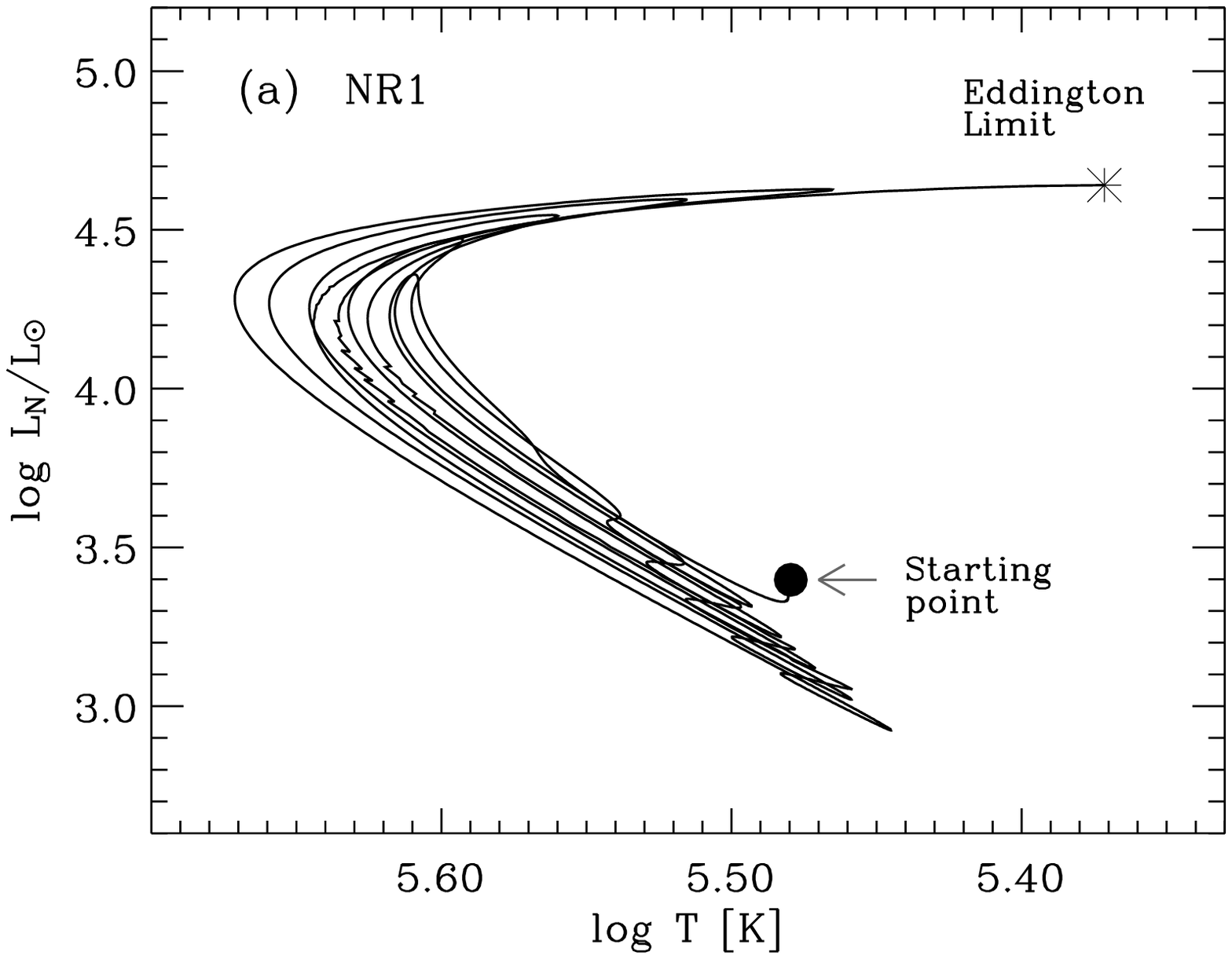}
\epsfxsize=\hsize
\epsffile{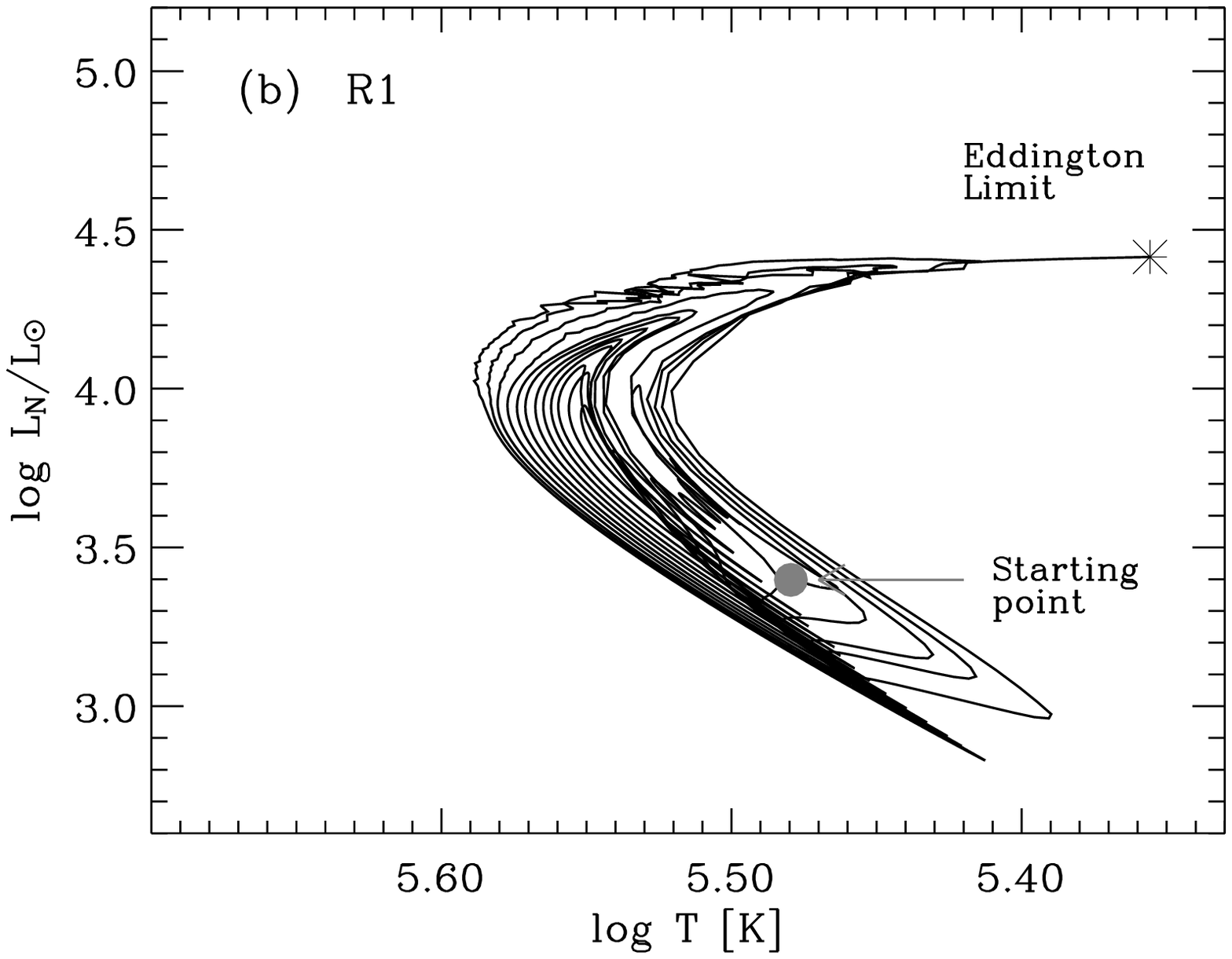}
\caption{Evolution of white dwarf 
in the sequences NR1 (a) and R1 (b) in the Hertzprung-Russell diagram.
The starting point
is marked by a filled circle. The ending point, where the surface luminosity
reaches the Eddington limit, is marked by an asterisk for each case.
The ruggedness which appears in some parts of the evolutionary tracks is due to the limited
number of digits in the numerical outputs. 
}\label{fig:HR}
\end{figure}

\begin{figure}[t]
\epsfxsize=\hsize
\epsffile{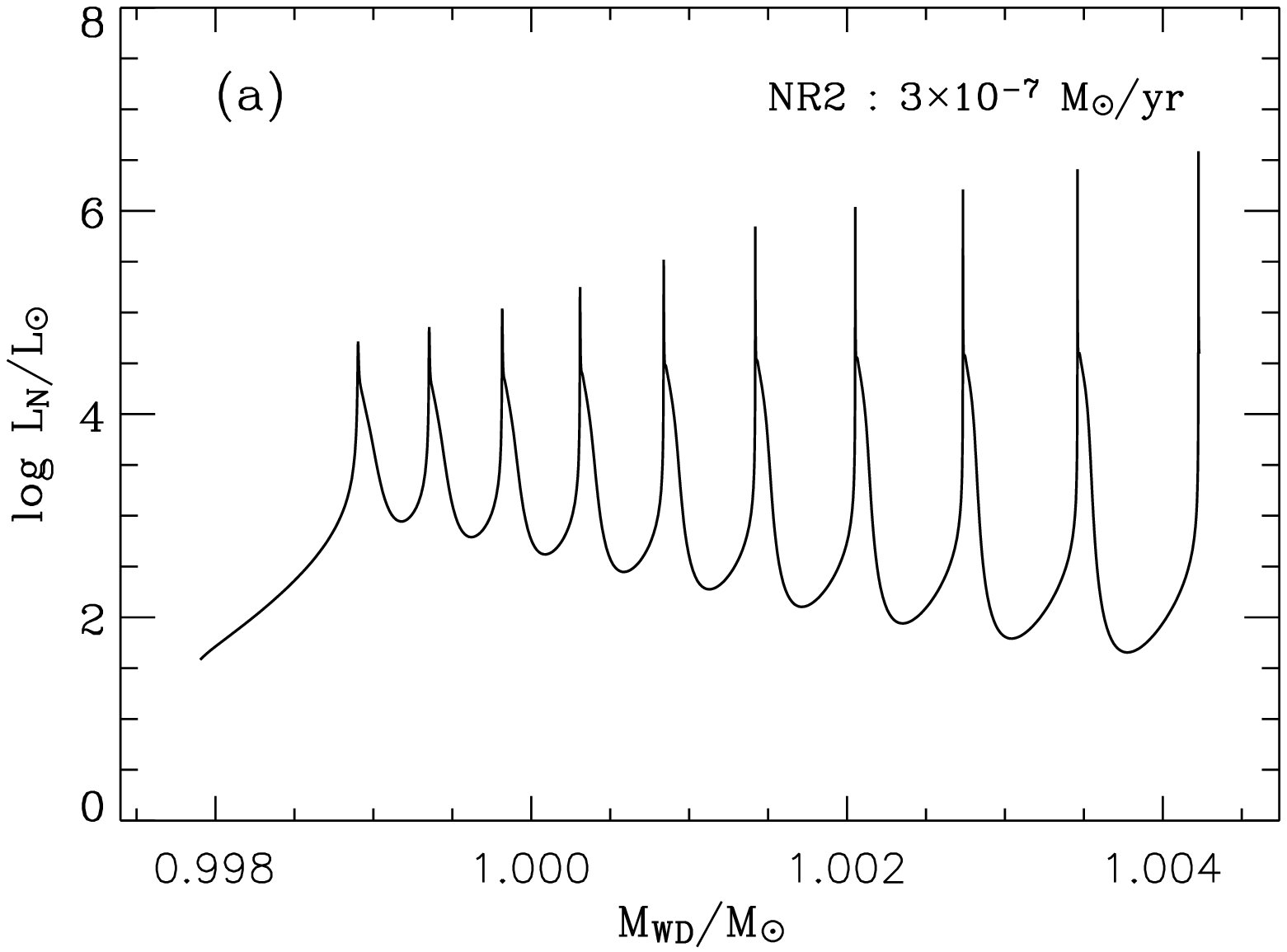}
\epsfxsize=\hsize
\epsffile{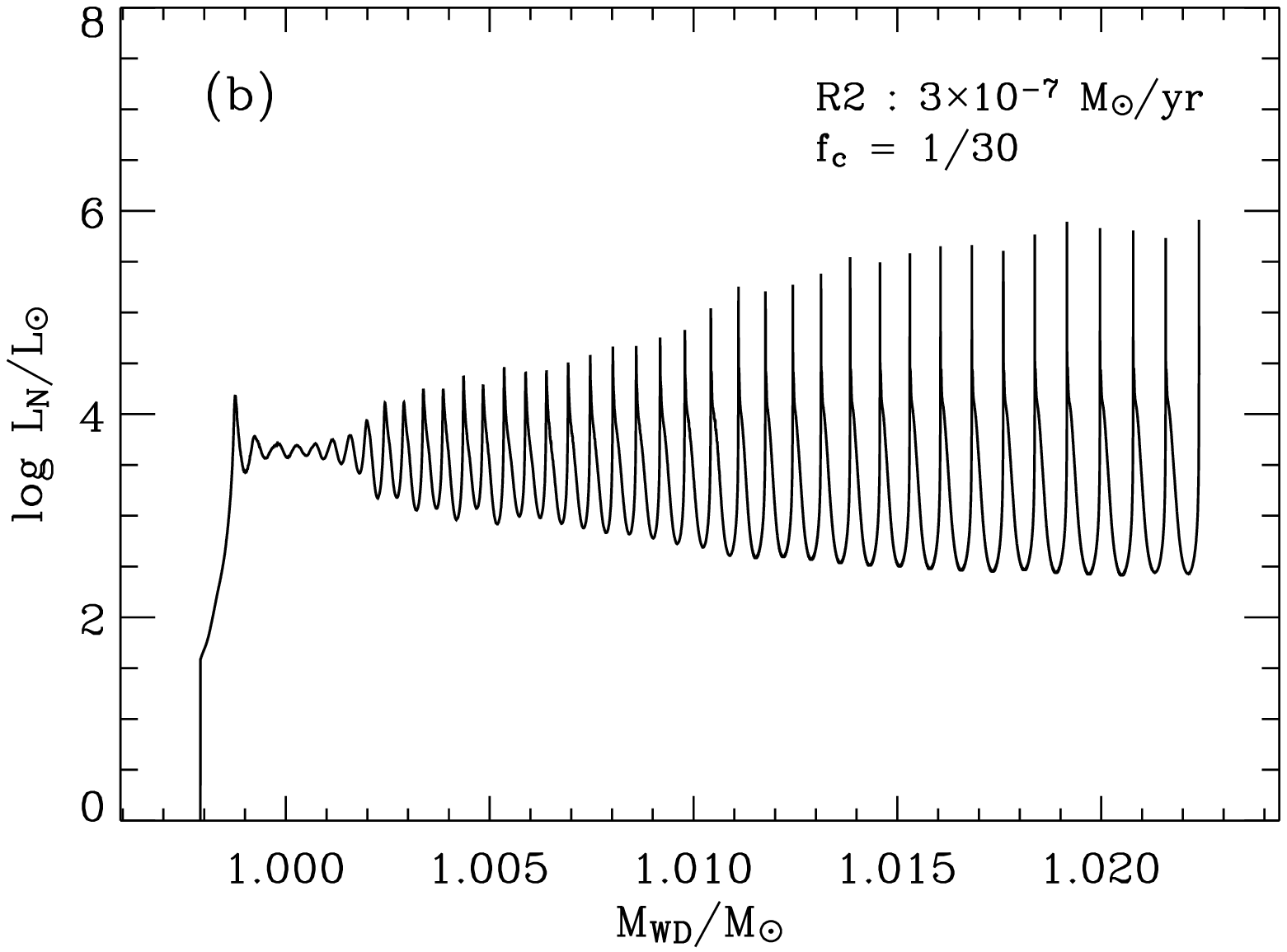}
\epsfxsize=\hsize
\epsffile{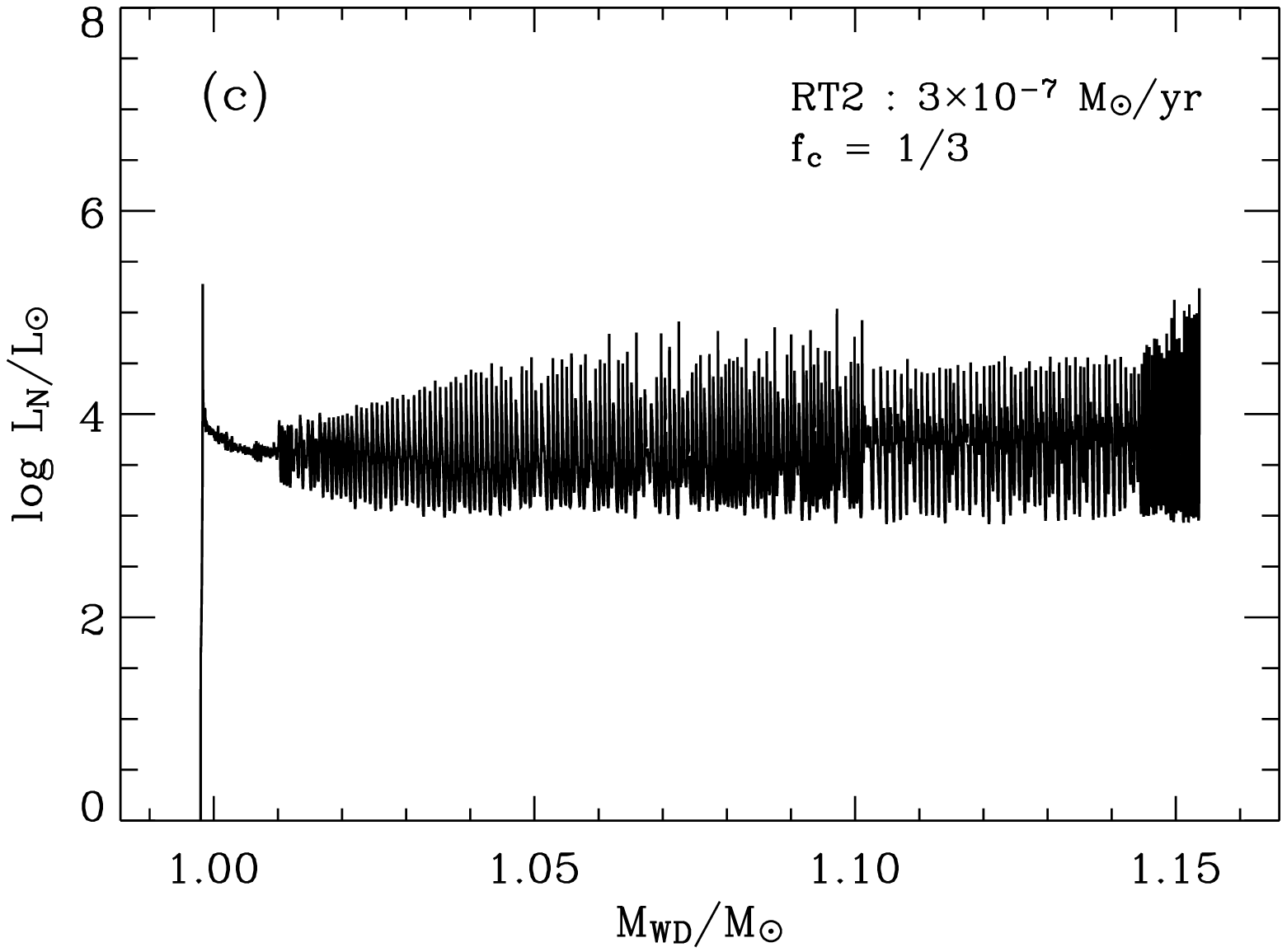}
\caption{
Same as in Fig.~\ref{fig:lhe1} but with \Mdot{} = \hemdotb{}~\msyr{} (NR2 (a), R2 (b) and RT2 (c)). 
}\label{fig:lhe2}
\end{figure}

The results of our calculations are summarized in Table~\ref{tab1}. 
For sequences NR1, NR2, NR3, R1 and R2, 
we followed the evolution until 
the helium shell flashes become strong enough to drive the surface luminosity
to the Eddington limit.
For the rest, calculations are stopped after a large number of models has been
calculated (cf. Tab.~\ref{tab1}).

\subsection{Helium shell burning and luminosity}\label{sect:heshell}

In Fig.~\ref{fig:lhe1}, the nuclear luminosity due to the helium burning in the sequences with
\Mdot{} = \hemdota{} \msyr{} (NR1, R1 \& RT1)
is shown as a function of the total mass of white dwarf.
In sequence NR1,
helium shell burning is not stable but undergoes
thermal pulses, which are driven by the thin shell instability 
(e.g. Schwarzshild \& H\"arm~\cite{Schwarzshild65}; Weigert~\cite{Weigert66}).
It is well known that this phenomenon occurs also in intermediate mass stars in
their thermally pulsating asymptotic giant branch (TP-AGB) phase 
(e.g. Iben \& Renzini~\cite{Iben83a}).
The thermal pulses in sequence NR1 become stronger as the white dwarf
accretes more matter (cf. Fujimoto \& Sugimoto~\cite{Fujimoto79}). Finally, 
the shell flash at the sixth pulse  drives the surface
luminosity to the Eddington limit when \Mwd{} $\simeq$ 1.002 \Msun{} (cf.
Fig.~\ref{fig:HR}), at which point we stop the calculation. 
Perhaps, the white dwarf would experience a radiation driven
optically thick outflow at this stage,
which would reduce the mass accumulation efficiency 
(Kato \& Hachisu~\cite{Kato99}). It is beyond the purpose of this paper to
consider the ensuing evolution in detail.

In the rotating sequence R1 with the same accretion rate,
thermal pulses become significantly weaker compared to the non-rotating case. 
For instance, the helium shell source reaches $\log L_{\rm N}/L_{\odot} \simeq 6.5$ 
in the non-rotating case
when the white dwarf has accreted  about 0.004 \Msun{}, 
while  the rotating sequence reaches only  $\log L_{\rm N}/L_{\odot} \simeq 4.0$ at the
same epoch (see Fig.~\ref{fig:lhe1}). 
As a consequence, the white dwarf can accrete more mass ($\Delta M_{\rm acc} \simeq$ 0.0096 \Msun)
than in sequence NR1 ($\Delta M_{\rm acc} \simeq$ 0.0039 \Msun) 
before the surface luminosity reaches the Eddington limit.
In the HR diagram (Fig.~\ref{fig:HR}), the rotating sequence is located in lower effective 
temperatures than 
the non-rotating sequence, since the white dwarf envelope is expanded by the centrifugal force.

The helium shell source becomes more
stable with stronger chemical mixing. Results of sequence RT1 in Fig.~\ref{fig:lhe1}c,   
where the chemical mixing is 10 times stronger than in sequence R1,
show that the helium is burned more or less steadily until \Mwd{} $\simeq$ 1.01 \Msun{}. 
More than 100 pulses  were followed thereafter,
but the surface luminosity still did not reach the Eddington limit.

The evolution of sequence NR2 (\Mdot{} = \hemdotb{} \msyr) shown in Fig.~\ref{fig:lhe2} 
is similar to that of NR1, only
the white dwarf reaches the Eddington limit somewhat later, i.e., when \Mwd{} $\simeq$ 1.004 \Msun{}.
The thermal pulses are weakened in 
the corresponding rotating sequence with $f_{\rm c} = 1/30$ (R2, Fig.~\ref{fig:lhe2}b),
and about 4 times more mass could be accreted in the white dwarf of sequence R2 before
reaching the Eddington limit.
With a stronger chemical mixing ($f_{\rm c} = 1/3$, RT2, Fig.~\ref{fig:lhe2}c), steady
helium shell burning continues until \Mwd{} $\simeq$ 1.01 \Msun{},
and weak thermal pulses with $\log L_{\rm N}/L_{\odot} \lsim 5.0$ follow. 
The evolution of this sequence was stopped when \Mwd{} $\simeq$ 1.15~\Msun{},
as the number of thermal pulses before reaching the Chandrasekhar mass 
would exceed $10^4$.

\begin{figure*}[!]
\epsfxsize=0.5\hsize
\epsffile{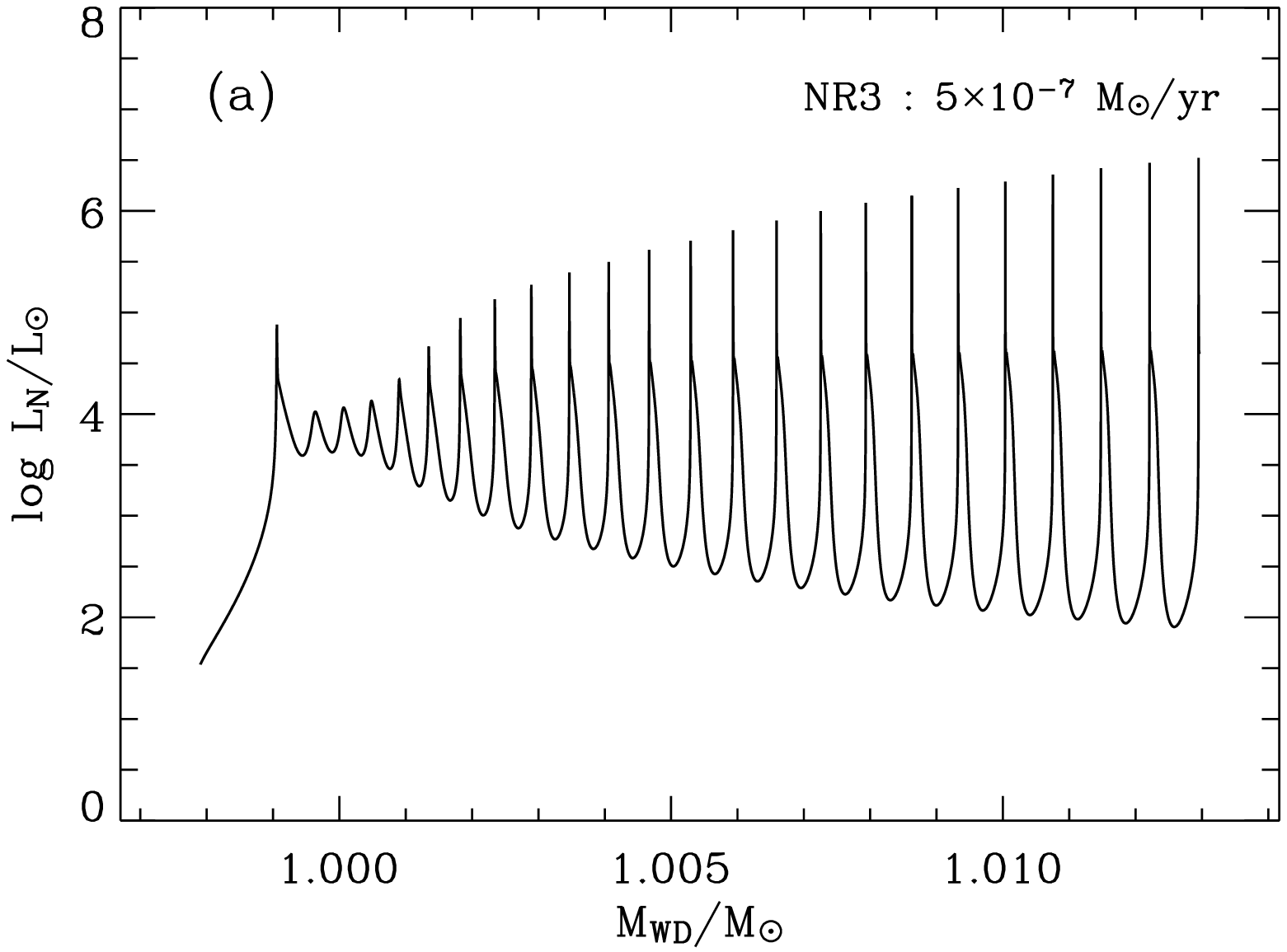}
\epsfxsize=0.5\hsize
\epsffile{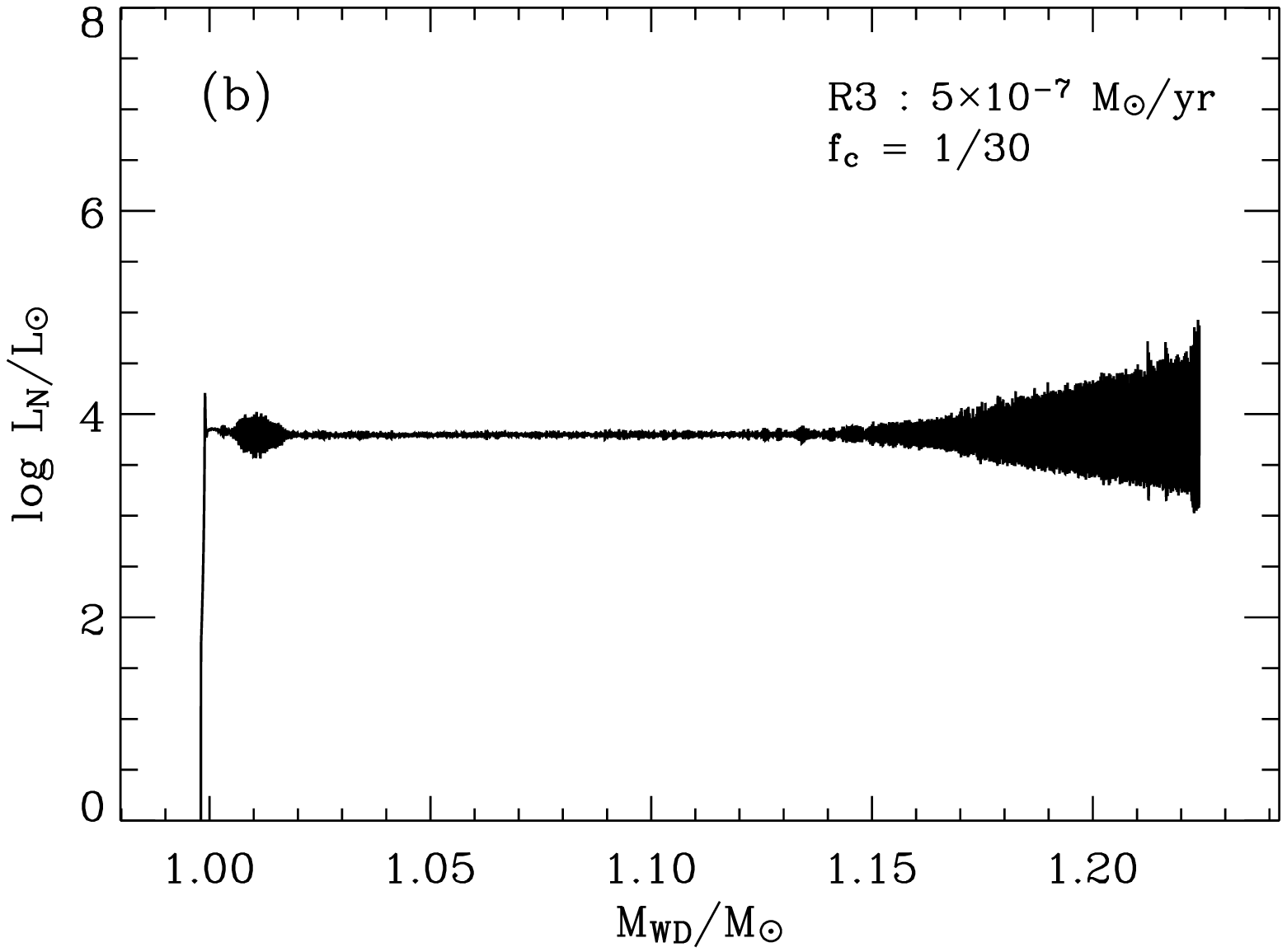}
\epsfxsize=0.5\hsize
\epsffile{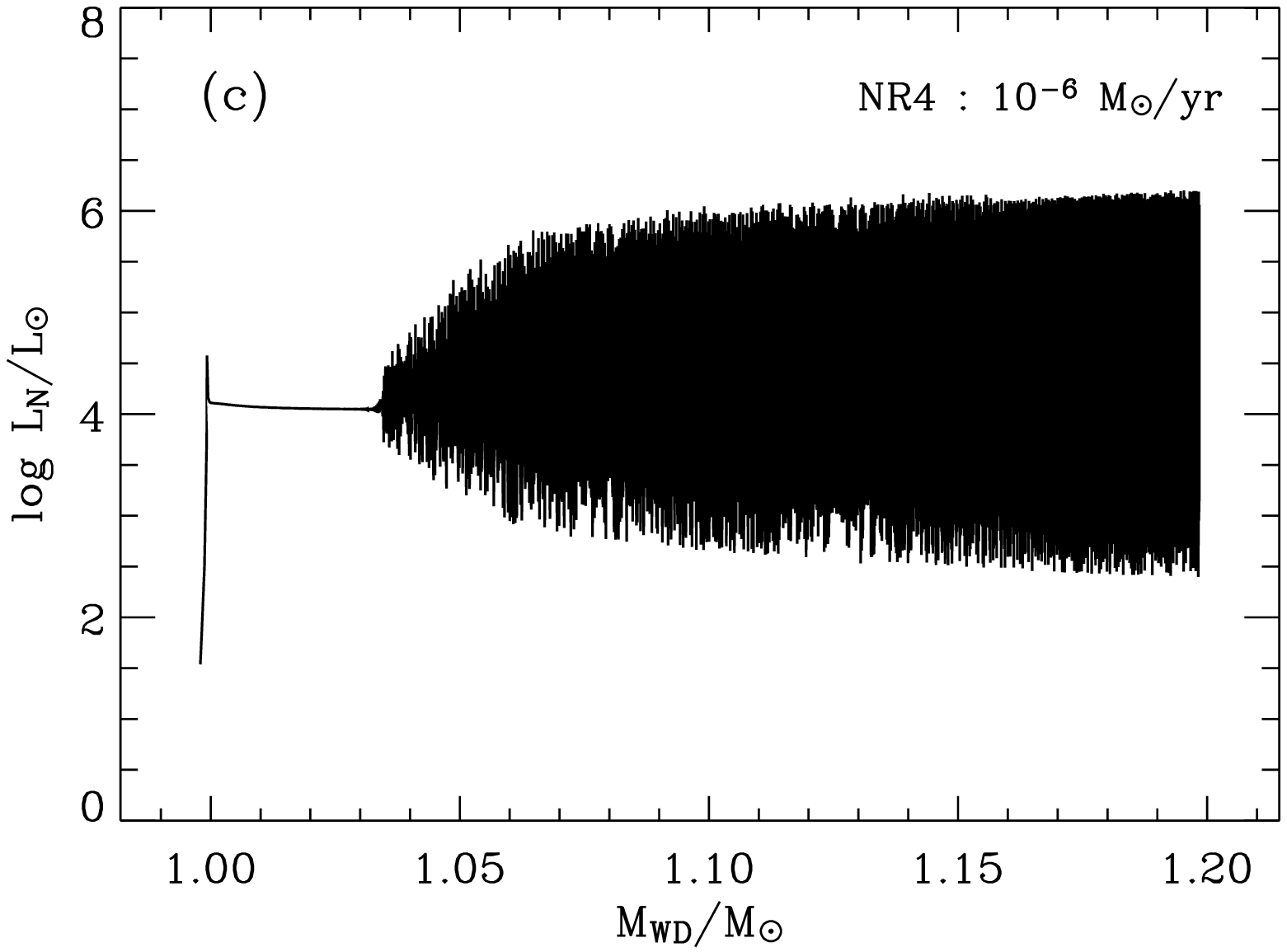}
\epsfxsize=0.5\hsize
\epsffile{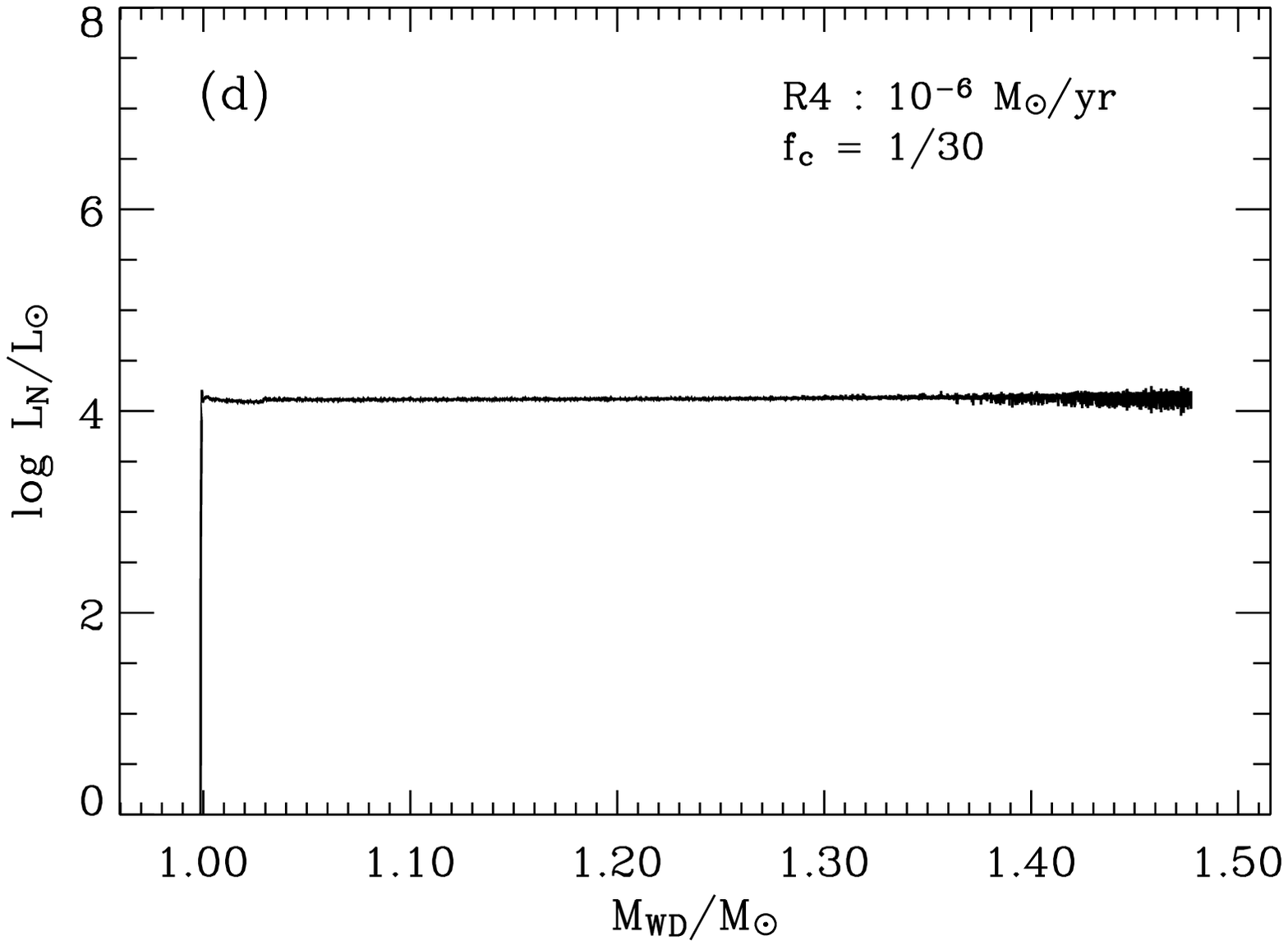}
\caption{Evolution of nuclear luminosity due to helium burning in the helium accreting white
dwarf models for \Mdot{} = \hemdotc{}~\msyr{} (NR3 (a) and R3 (b)) 
and \Mdot{} = \hemdotd{}~\msyr{} (NR4 (c) and R4 (d)).
The abscissa denotes the total mass of white dwarf, which serves as a measure of time.
Note that what appears to look like noise in the plots for R3 and NR4
corresponds to many successive thermal pulses each of which being as well resolved in time
(i.e., by roughly $10^3$ time steps) as those in the plots in Fig.~\ref{fig:lhe1} and~\ref{fig:lhe2}.
}\label{fig:lhe3}
\end{figure*}

The stabilizing effect of rotation
appears more prominently at \Mdot{} = \hemdotc{} \msyr. 
In the non-rotating case (NR3),
the helium shell burning is still unstable even with such a high
accretion rate, as shown in Fig.~\ref{fig:lhe3}a, and
the white dwarf reaches the Eddington limit when \Mwd{} $\simeq$ 1.013 \Msun.
In the corresponding rotating sequence (R3), very weak thermal pulses
appear when \Mwd{} $\simeq$ 1.005~\Msun{},
but are suppressed when \Mwd{}~$\gsim$~1.02~\Msun{}.
Steady helium shell burning continues until \Mwd{}~$\simeq$~1.14~\Msun{},
from which the shell source becomes again unstable. 
Note that the thermal pulses shown
in Fig.\ref{fig:lhe3}b look like noise
but are actually many successive thermal pulses, each of which 
is  resolved by about $10^3$ models.

With an accretion rate of $\dot{M}=10^{-6}$ \msyr, 
the non-rotating white dwarf undergoes unstable helium shell burning
after \Mwd{} $\simeq1.03$ \Msun{} (NR4, Fig.~\ref{fig:lhe3}c).
We have followed the evolution until \Mwd{} = 1.2 \Msun{} 
and find that thermal pulses do not become strong enough to lead
the white dwarf to the Eddington limit. 
On the other hand, in the rotating sequence R4 (Fig.~\ref{fig:lhe3}d), 
the white dwarf
can grow up to \Mwd{} $\simeq1.48$\Msun{} without encountering significant
thermal pulses.

\begin{figure}[!]
\epsfxsize=\hsize
\epsffile{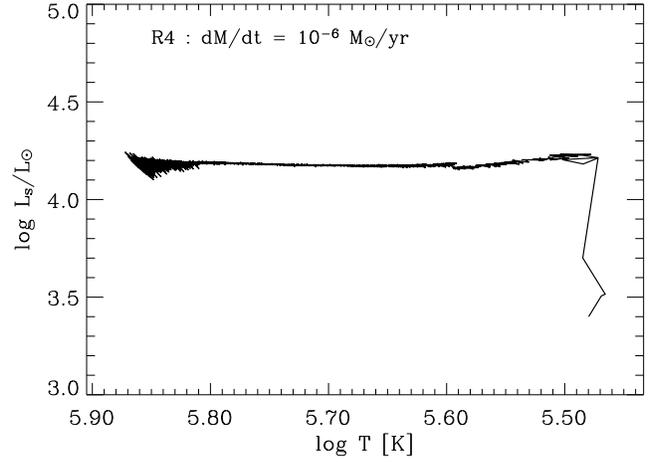}
\caption{Evolution of white dwarf 
in sequence R4 in the Hertzsprung-Russell diagram.
The region with $\log L_{\rm s}/L_{\odot} = 3.4\dots4.2$
and $\log T = 5.46\dots5.48$ represents the epoch from the onset of
helium accretion to the first mild helium shell flash (cf. Fig.~\ref{fig:lhe3}d).
}\label{fig:hr2}
\end{figure}

The evolution of the white dwarf of sequence R4 in HR diagram is shown 
in Fig.~\ref{fig:hr2}.
The surface luminosity 
remains nearly constant at $\log L_{\rm s}/L_{\odot} \simeq 4.2$, 
while the surface temperature increases from $\log T = 5.46$ to $\log T = 5.88$,
with which the white dwarf will appear as a super-soft X-ray source 
(e.g. Kahabka \& van den Heuvel~\cite{Kahabka97}; Greiner~\cite{Greiner00}).

Results presented in this section  show  that
the helium shell source becomes more stable with rotation, and
that the stronger the chemical mixing the more stable 
the shell burning becomes. We discuss the physical reasons
for this behavior in Sect.~\ref{sect:stability}.

\subsection{White dwarf spin and angular momentum transport}\label{sect:spin}

\begin{figure}[!]
\epsfxsize=\hsize
\epsffile{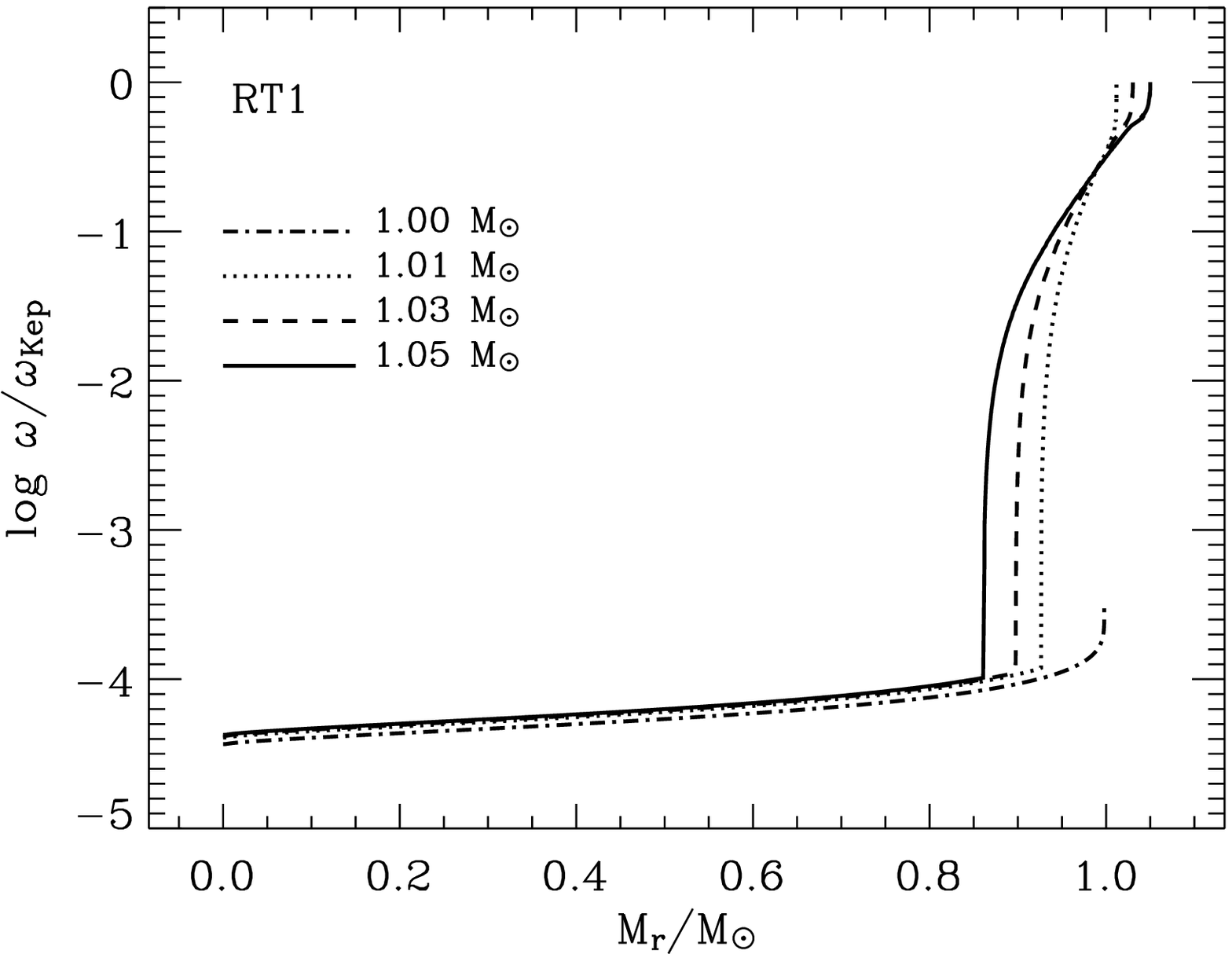}
\epsfxsize=\hsize
\epsffile{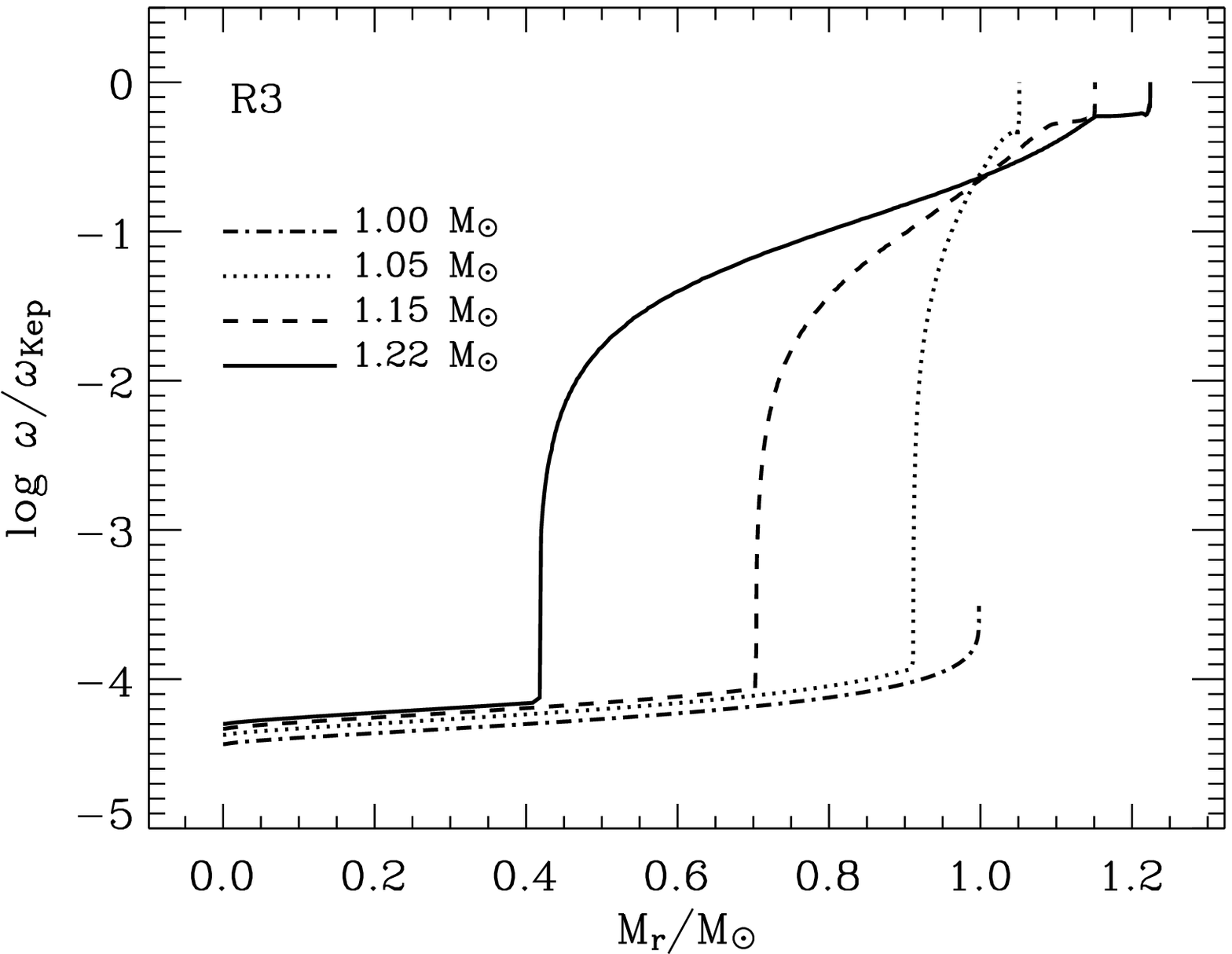}
\epsfxsize=\hsize
\epsffile{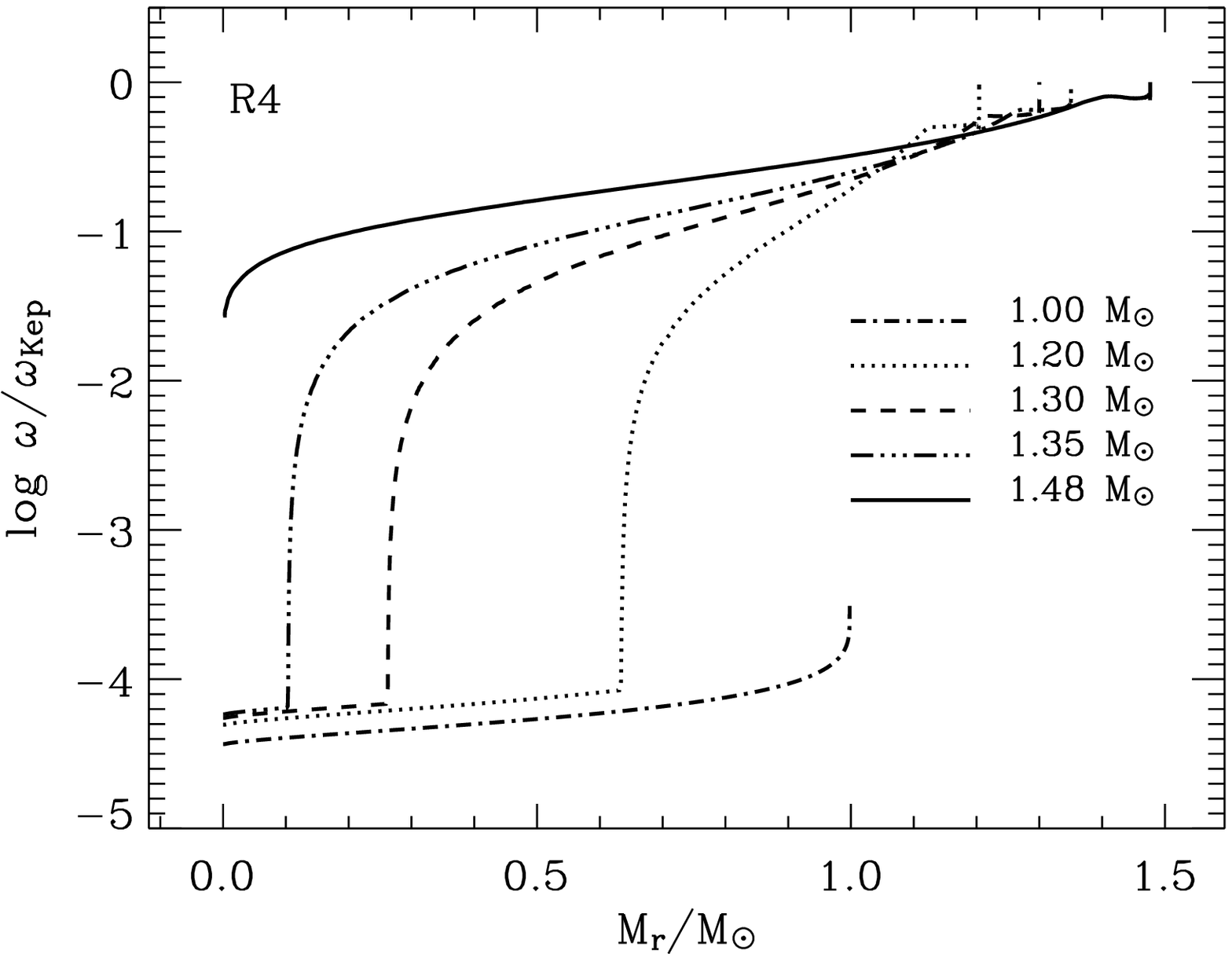}
\caption{Spin velocity at different white dwarf masses, 
normalized to the local Keplerian value, as function
of the mass coordinate for RT1, R3 and R4.
}\label{fig:spin}
\end{figure}

According to Yoon \& Langer (\cite{Yoon04a}), who considered accretion of 
carbon-oxygen rich matter, 
non-magnetic accreting white dwarfs with \Mdot{} $\gsim 10^{-7}$ \msyr{} 
show the following features in relation to the angular momentum
redistribution.
a) Eddington-Sweet circulations, 
the secular shear instability, and the GSF instability are
dominating the angular momentum redistribution in the non-degenerate outer envelope,
while the dynamical shear instability is the most important in the degenerate core. 
b) The degree of differential rotation in the degenerate core 
remains strong throughout the evolution, with a shear strength
near the threshold value for the onset of the dynamical shear instability.
c) The resulting differential rotation changes
the white dwarf structure such that the center
does not reach carbon ignition 
even when the white dwarf reaches the Chandrasekhar limit ($\sim$ 1.4 \Msun).
d) More than 60 \%{} of the angular momentum of the accreted matter
is rejected by the condition that the white dwarf should not gain
angular momentum when its surface rotates critically (Eq.~\ref{eq:1}). 

It is found that all these features remain the same
in the present study, where helium accretion is considered.
Fig.~\ref{fig:spin} shows angular velocity profiles throughout white
dwarf models of sequence RT1, R3 and R4, at various times, 
where the angular velocity is given in units 
of the local Keplerian value (i.e., $\omega_{\rm Kep}=\sqrt{GM_r/r^3}$).
Note that the spin rate remains well below the local Keplerian value
throughout the whole star in all cases.

The white dwarf model with \Mwd{} $\simeq$ 1.4 \Msun{} in 
sequence R4 has a central density of 3.54 $\times 10^8~{\rm g/cm^3}$,
which is still far from the carbon ignition density.
(cf. Ostriker \& Bodenheimer~\cite{Ostriker68a}; Durisen~\cite{Durisen75b}; Durisen \& Imamura~\cite{Durisen81}).
As shown in the mentioned papers, a differentially rotating
white dwarf can be dynamically stable up to $\sim$ 4.0 \Msun.
As discussed in Yoon \& Langer (\cite{Yoon04a}), this suggests that the secular
instability to the gravitational wave radiation reaction, which
is often called as the 'CFS instability' (Chandrasekhar~\cite{Chandrasekhar70}; Friedman \& Schutz~\cite{Friedman78}),
might be important for the final fate of accreting white dwarfs 
at near or above the Chandrasekhar limit. 
The ratio of the rotational energy to the gravitational energy (\TW), 
which determines the growth time scale of the CFS instability via
the bar-mode or the $r$-mode (see Yoon \& Langer \cite{Yoon04a}), remains smaller than 0.1
in the white dwarf models of sequence R4, and the last 
model (\Mwd{} = 1.48 \Msun) has \TW{} = 0.06. 
Since \TW{} $\gsim$ 0.1  is required in white dwarfs with strong differential rotation 
for the onset of the bar-mode instability (Imamura et al.~\cite{Imamura95}), 
only the $r$-mode may be relevant at this point.
The expected growth time of the $r$-mode instability
in this model is about 1.1$\times 10^5$ yr 
(see Yoon \& Langer~\cite{Yoon04a}; cf. Lindblom \cite{Lindblom99}). 
We refer to Yoon \& Langer (\cite{Yoon04a}) 
for a detailed discussion of the final fate  of rapidly
rotating massive white dwarfs.

\section{Chemical mixing and nucleosynthesis}\label{sect:mix}

\begin{figure}[!]
\epsfxsize=\hsize
\epsffile{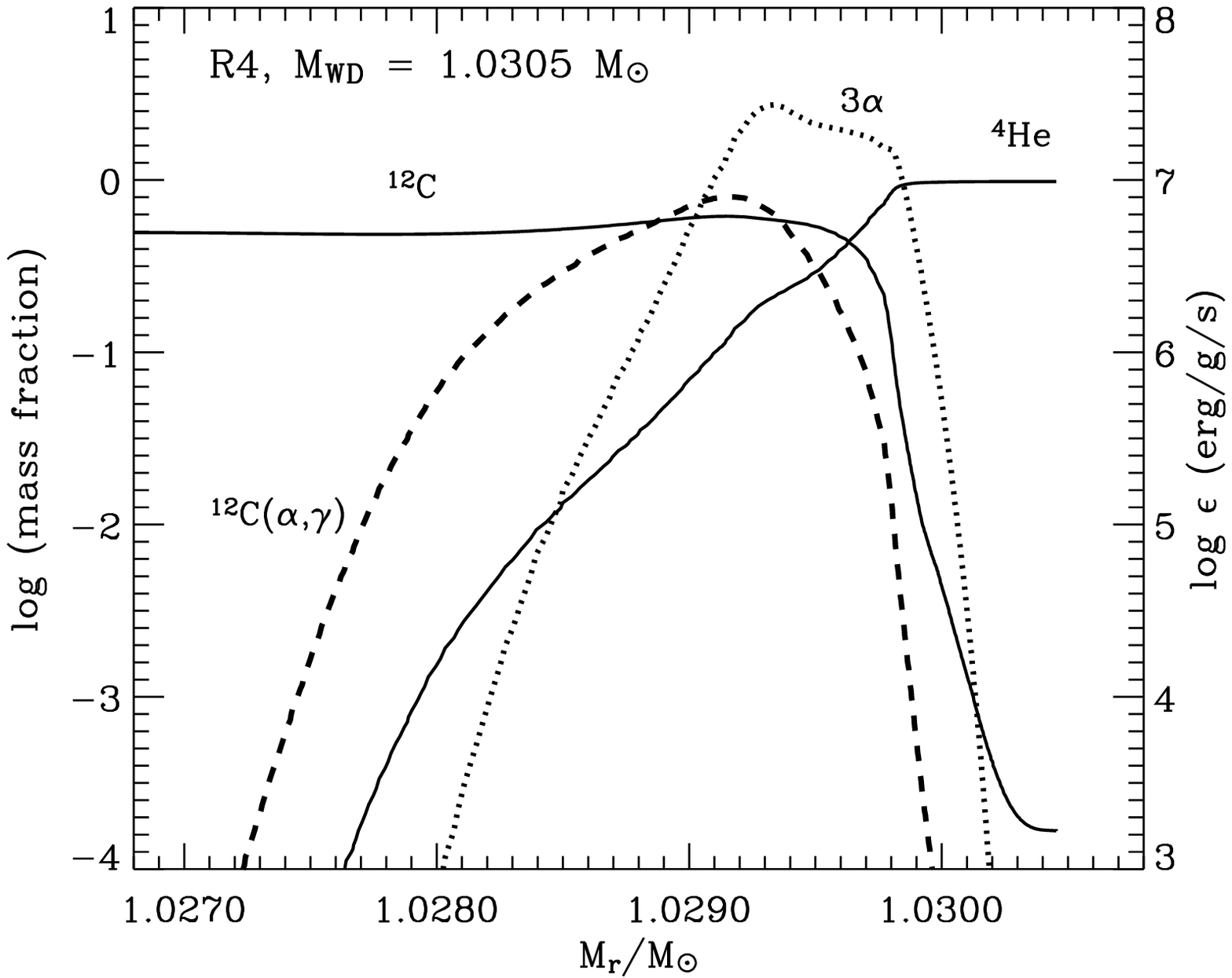}
\epsfxsize=\hsize
\epsffile{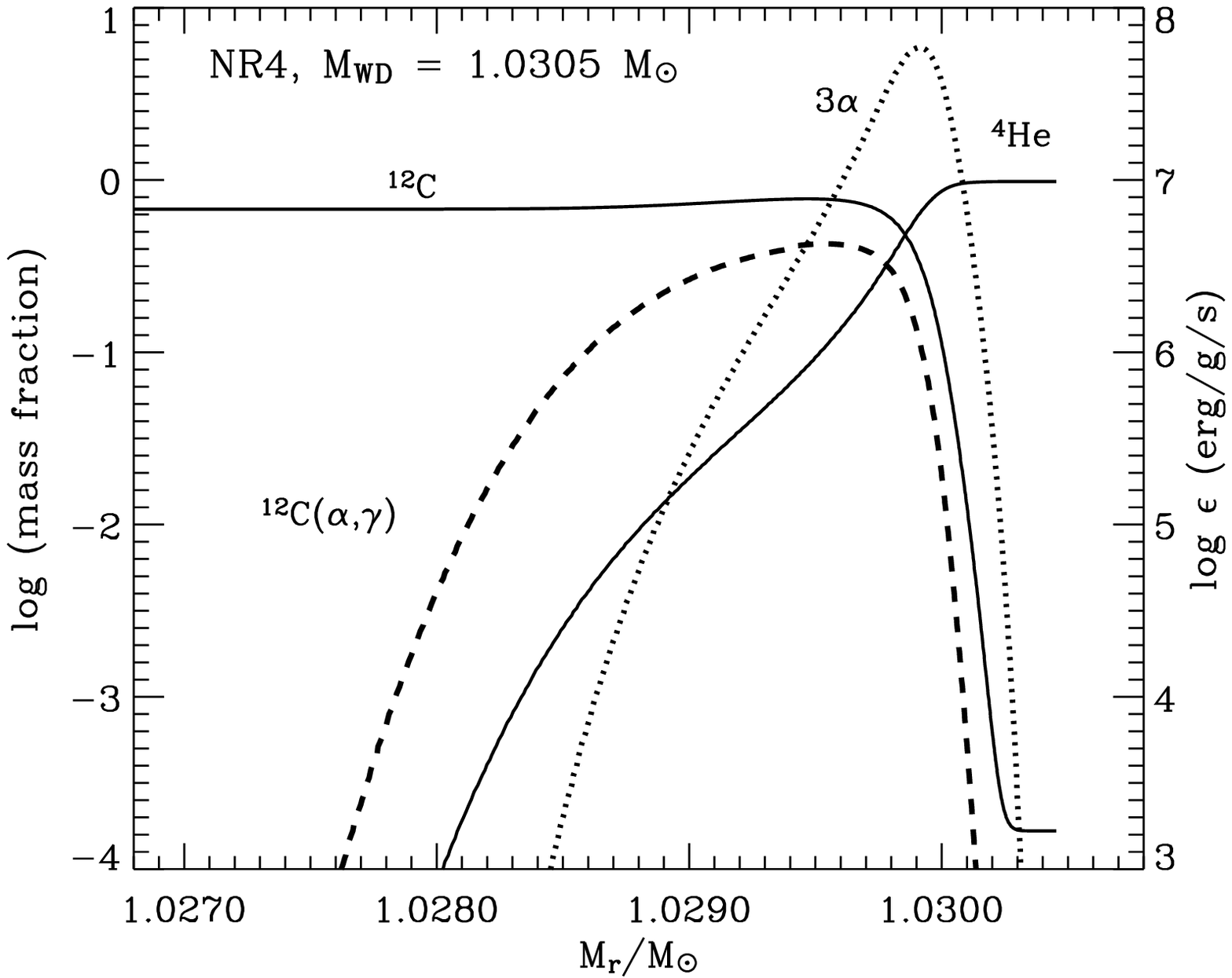}
\caption{Logarithm of the mass fractions of helium and carbon as function 
of the mass coordinate (solid line), in the white dwarf models of sequences NR4 
(upper panel) and R4 (lower panel) at a time when
 \Mwd{} = 1.0305 \Msun. The dotted and dashed lines denote
the corresponding energy generation rate (right scale)
for the $3\alpha$ and $^{12}\mathrm{C}(\alpha,\gamma)$ 
reactions respectively.
}\label{fig:chem}
\end{figure}

\begin{figure}[!h]
\epsfxsize=\hsize
\epsffile{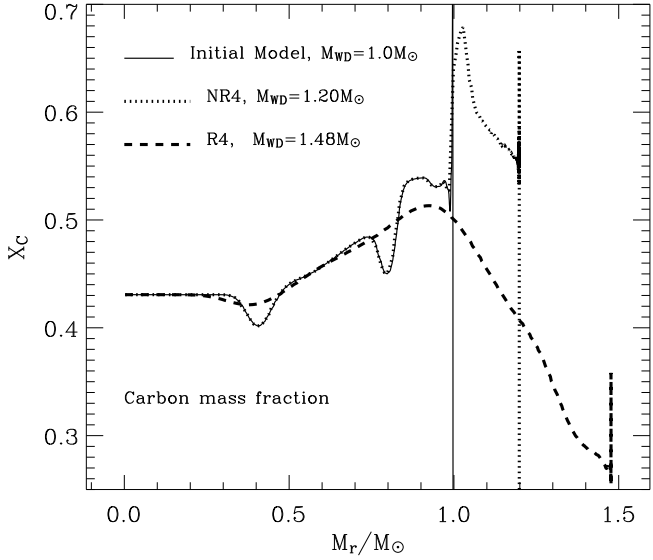}
\caption{Carbon mass fraction as a function of the mass coordinate 
for the last computed models of sequences R4 (dashed line) and NR4
(dotted line). The thin solid line corresponds to the initial model.
Note that the mass of the helium envelope is
only of the order of $10^{-4}$ \Msun, which is not resolved
in this figure.  The sharp peak in the carbon abundance near
the surface at each white dwarf model is due to the active
$3\alpha$ reaction at the bottom of the helium envelope
(cf. Fig.~\ref{fig:chem}).
}\label{fig:xc}
\end{figure}

The mixing of $\alpha$-particles into the CO-core and of carbon into
the helium envelope has consequences for the helium
burning shell source. Here, we discuss the resulting changes in
chemical structure of the white dwarf, while in the next section
we analyze how rotation and rotationally induced chemical mixing
affects the stability properties of the shell source.

Fig.~\ref{fig:chem} shows the chemical structure 
and the nuclear energy generation rate
in the helium  burning shell in the white dwarf 
models of sequences R4 and NR4 when $M_{\rm WD}=1.0305$ \Msun. 
The helium layer in the rotating model extends farther into the CO core,
compared to the non-rotating case,  due to the rotationally induced
chemical mixing. This renders the \rche{} reaction more active in the rotating model,
with two main consequences.

Firstly, the contribution of the  \rche{} reaction to
the total energy production is considerably increased. 
In the given models in Fig.~\ref{fig:chem}, 
the rotating model gives $\log L_{3\alpha}/L_{\odot} = 3.938$
and $\log L_{^{12}\mathrm{C}(\alpha,\gamma)}/L_{\odot} = 3.562$, 
while the non-rotating model gives $\log L_{3\alpha}/L_{\odot} = 3.963$
and $\log L_{^{12}\mathrm{C}(\alpha,\gamma)}/L_{\odot} = 3.338$, 
where $L_{3\alpha}$ and $L_{^{12}\mathrm{C}(\alpha,\gamma)}$
are the integrated nuclear energy generation rate over the shell source,
due to the triple alpha reaction and the \rche{} reaction, respectively.
The \rtrip{} reaction is weaker in the rotating model than 
in the non-rotating one
due to a lower density  
in the helium envelope 
(cf. Fig.~\ref{fig:stab})
and  due to a smaller mass fraction of helium at the peak of the nuclear energy generation rate
caused by rotationally induced mixing (Fig.~\ref{fig:chem}). 
However, the increase of the energy generation by the \rche{} reaction
is more important on the whole, 
leading to a higher nuclear luminosity in the rotating model 
($\log L_{\rm N}/{\rm L_{\odot}} = 4.094$)
compared to the corresponding non-rotating case ($\log L_{\rm N}/{\rm L_{\odot}} = 4.060$),
for the models given in Fig.~\ref{fig:chem}.
Furthermore, the shell source is hotter in the rotating model
[$T_{\rm shell}/(10^8~{\rm K})=2.16$] 
than in the comparable non-rotating model [$T_{\rm shell}/(10^8~{\rm K}) = 2.03$],
which affects the stability of the shell source as 
discussed in Sect.~\ref{sect:stability}.

Secondly, the enhanced $^{12}\mathrm{C}(\alpha,\gamma)$-efficiency results in
an increased oxygen abundance in the ashes of the helium burning shell,
at the expense of carbon. Fig.~\ref{fig:xc} shows the carbon mass fraction
throughout the white dwarf for the last computed models of sequences~NR3 and~R3. 
In the non-rotating model, the carbon mass fraction produced by the shell
source reaches almost $X_{\rm C}\simeq 0.68$ at a white dwarf mass of
$1.03$ \Msun. At this time, the helium shell source becomes unstable
and develops thermal pulses. As the layers at and above the shell source
become convectively unstable during each pulse, the resulting carbon abundance
is reduced. The convective mixing becomes
more and more significant as thermal pulses become stronger (cf. Fig.~\ref{fig:lhe3}c),
which is reflected in the drop of $X_{\rm C}$ from about 0.68
to 0.55 in the range 1.03~\Msun{} $\lsim M_{\rm r} \lsim 1.20$~\Msun{} in Fig.~\ref{fig:xc}.
A comparison with the rotating model shows that the carbon mass fraction is
mostly below 0.5, even though the rotating model does not suffer from thermal
pulses for most of the time. 
While the average carbon mass fraction in the accreted layer is 
0.58 in the non-rotating model, it is 0.45 in the 
layer with $1.0 < M_r/{\rm M_{\odot}} < 1.2$ of the
rotating model. The mean value of $X_{\rm C}$ throughout the
accreted layer in the rotating model (i.e., $1.0 < M_r/{\rm M_{\odot}} < 1.48$) 
is 0.38.

The carbon and oxygen mass fractions in the accreted envelope of
a  SN~Ia progenitor are important from the point of view of nucleosynthesis
and supernova spectroscopy, and may even affect  
the peak brightness of the supernova explosion 
(cf. Umeda et al.~\cite{Umeda99}; H\"oflich et al.~\cite{Hoeflich00}; Dom\'inguez et al.~\cite{Dominguez01}).

\section{Stability of helium shell burning}\label{sect:stability}
 
\begin{figure*}[!ht]
\begin{center}
\epsfxsize=0.7\hsize
\epsffile{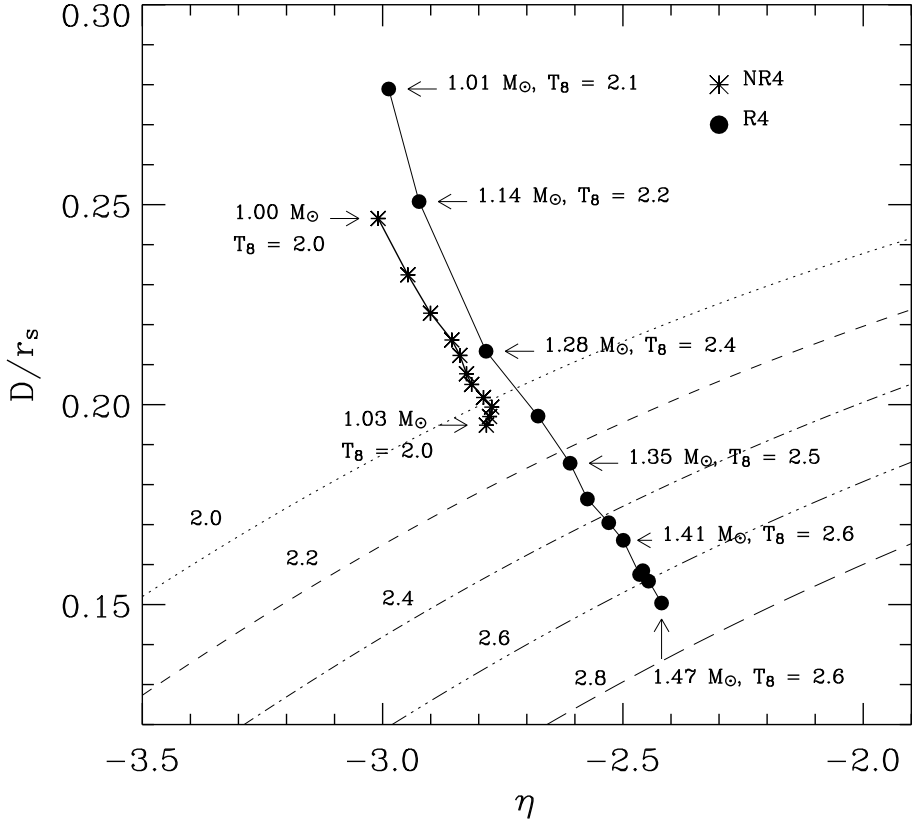}
\end{center}
\caption{Evolution of the helium shell source in the sequences NR4 and 
R4 during stable shell burning phase, in the plane spanned by  
the degeneracy parameter $\eta$ and the relative thickness of shell source $D/r_{\rm s}$.
Here $D$ denotes the thickness of shell source and $r_{\rm s}$
its radius. Each asterisk and black dot represents one selected stellar model
in sequence NR4 and R4, respectively. 
The white dwarf mass in solar masses and the temperature of the shell source
in the unit of $10^8$ K are indicated for various models.
The thin lines separate the stable and unstable regime, 
for $T/(10^8 ~{\rm K}) = $ 2.0  (dotted line), 2.2 (short dashed line), 2.4 (dotted-dashed line),
2.6 (three dotted - dashed line) and 2.8 (long dashed line)
according to the analytic stability criterion of Yoon, Langer \& van der Sluys (\cite{Yoon04d}).
I.e., helium shell burning is predicted to be stable (unstable)  above (below)  the lines,
while thermal pulses become significant from $M_{\rm WD}\simeq 1.03 $ \Msun{}
in sequence~NR4, and weak thermal pulses appear 
from $M_{\rm WD} \simeq 1.41 $ \Msun{} in sequence R4. 
}\label{fig:stab}
\end{figure*}

Our results show that helium shell burning is more stable with rotation,
compared to the non-rotating case (Figs.~\ref{fig:lhe1},~\ref{fig:lhe2},~\ref{fig:lhe3}). 
The stability of a helium shell source is mainly determined
by three factors: geometrical thickness, degree of degeneracy
and temperature of the shell source. 
In other words, a helium shell source is more stable if it is thicker, 
less degenerate and hotter. While a general and quantitative stability criterion 
for shell sources is worked out in a separate paper (Yoon, Langer \& van der 
Sluys~\cite{Yoon04d}), we discuss the qualitative effects of rotation on the
shell source stability of the models presented in Sect.~3 in the following.

Fig.~\ref{fig:stab} shows the evolution of the helium shell 
sources of sequences NR4 and R4
in the plane of the degeneracy parameter $\eta$ ($:=\psi/kT$, e.g. Clayton~\cite{Clayton68})
and the relative thickness of the shell source $D/r_{\rm s}$, where
$D$ and $r_{\rm s}$ are the thickness and the radius of the shell source, respectively.
Here, the energy generation rate
weighted mean over the shell source is used for the degeneracy parameter.
The thickness of the shell source is defined such that
the energy generation rate at each boundary is $2\times10^{-3}$ times its peak value.
Fig.~\ref{fig:stab} illustrates the evolution toward instability for
the example of sequences NR4 and~R4.
In sequence~NR4, the mean temperature in the shell source
remains nearly constant, around $T\simeq 2.02\times 10^8~{\rm K}$ during
the stable shell burning phase. 
Fig.~\ref{fig:stab} shows the border between stability
and instability for $T=2\times10^8$ K, as obtained by Yoon, Langer \& van der 
Sluys (\cite{Yoon04d}), as dotted line.
The shell source becomes more degenerate and thinner
as the white dwarf mass grows, and finally enters 
the unstable regime when \Mwd{}~$\simeq1.03$~\Msun, resulting
in the onset of thermal pulses as shown in Fig.~\ref{fig:lhe3}c.

As already discussed in Sect.~\ref{sect:mix}, the shell source in the rotating models
extends over a wider mass range due to rotationally induced chemical mixing. I.e.,
significant amounts of $\alpha$-particles are mixed into the CO core and participate
in the \rche{} reaction. The presence of the centrifugal force
lowers the density, and were it not for the effect of mixing, the shell source temperature
would decrease with rotation. However, due to the enhanced \rche{} reaction,
temperature and nuclear energy production are increased, despite the lower density.
In summary, rotation changes all three physical properties
which affect the stability --- thickness, degeneracy and temperature --- such that
stability becomes more likely. 
Fig.~\ref{fig:stab} shows that this effect is quite significant.
While the evolutionary tracks of sequences~NR4 and~R4 are close to each other in the
shell thickness-degeneracy plane, {\em at a given mass} the shell thickness
is much larger and the degeneracy much lower in the rotating models. 
For example, at \Mwd{} = 1.01 \Msun{}, the non-rotating model
gives $\eta \simeq  -2.9$ and $D/r_{\rm s} \simeq 0.225 $ and
the corresponding rotating model gives $\eta \simeq -3.0$ and 
$D/r_{\rm s} \simeq 0.281$. 
Furthermore,
we have $\eta \simeq -2.8$ and $D/r_{\rm s} \simeq 0.19$ at \Mwd{} = 1.03 \Msun{}, from which
thermal pulses become significant in the non-rotating sequence, 
while $\eta \simeq -2.9$ and $D/r_{\rm s}\simeq0.25$ even 
when \Mwd{} is as large as 1.14 \Msun{} in the rotating model (Fig.~\ref{fig:stab}). 

Additionally, the instability border which is given by
the thin lines in Fig.~\ref{fig:stab} is located lower 
at a given mass in the rotating case,
due to the higher shell source temperature in the rotating models, 
which leads to a later onset of the instability even for 
a given shell source thickness and degeneracy. 
For instance, the shell source temperature in the rotating model
when \Mwd{} = 1.01~\Msun{} is about $ 2.1 \times 10^8$~K, which is higher
than in the non-rotating models, where $T_{\rm shell}$ remains at $ \sim 2.0 \times 10^8$~K
during the steady shell burning phase. \
The last analyzed model in sequence NR4 is already in the unstable regime 
(i.e., below the dotted line; $D/r_{\rm s} \simeq 0.19$) 
and thermal pulses become significant right after as expected from the stability criterion. 
On the other hand, in sequence R4, the shell source temperature increases up to  $ 2.6 \times 10^8$~K
when \Mwd{} = 1.41~\Msun{}, which explains why the shell source
is stable until \Mwd{} $\simeq$ 1.41~\Msun{} even though 
$D/r_{\rm s}$ at this moment is smaller than in the last analyzed model of NR4.
As shown in Fig.~\ref{fig:lhe3}d (see also Fig.~\ref{fig:hr2}), 
the ensuing thermal pulses which appear when \Mwd{} $\gsim$ 1.41~\Msun{}  in sequence R4
remain very weak until the last model 
(\Mwd{} = 1.48 \Msun{} with $T_{\rm shell} \simeq 2.6 \times 10^8 $~K), 
which is still near the stability border line for $T=2.6\times10^8$~K 
in Fig.~\ref{fig:stab}.

In summary, rotation stabilizes the instability of the 
helium shell burning in our models significantly. The shell sources in
realistic white dwarfs, however, may be even more stable than 
our models predict, due to two reasons. 
Firstly, as outlined in Sect.~\ref{sect:method}, our method
forces us to limit the structural effects of rotation to that 
obtained at about 60\% of critical rotation. This means that we underestimate
the centrifugal force in the layers which contain the shell source.
Secondly, our models are one-dimensional, which restricts
the pressure response due to any perturbation to the radial direction ---
which is generally not true in the realistic situation. In a rapidly rotating
star, the pressure can also respond to a perturbation 
horizontally, unless the shell source is located at the poles.
We believe that this multi-dimensional effects can stabilize the shell source further.

\section{Final discussion}

We have performed numerical simulations of helium accreting CO white dwarfs 
with an initial mass of 0.998 \Msun{} and 4 different
mass accretion rates (2, 3, 5 and 10$\times10^{-7}$ \msyr),
considering spin-up by mass accretion.
The main feature of the spin-up is 
the  angular momentum transport from the accreted matter into the white dwarf, 
which creates differential rotation in the white dwarf interior.
This induces transport of chemical species from the accreted matter into the CO core, 
through the shear instability,  Eddington-Sweet circulations, and the GSF instability.

It is shown that the helium burning layer 
becomes spatially more extended compared to the non-rotating case, 
since the accreted $\alpha$-particles
are mixed into the carbon enriched layer,  participating actively 
in the \rche{} reaction (Sect.~\ref{sect:mix}).
The carbon-to-oxygen ratio is significantly reduced
in the accreted layer. 
The centrifugal force lowers density in the shell source and degeneracy
is significantly lifted.
The total energy generation in the helium shell source increases  with rotation,
due to the increased contribution from the \rche{} reaction, rendering
the helium shell source as hot as  in the non-rotating cases, despite
the decreased density.
All these effects contribute to stabilizing the helium shell burning 
significantly (Sect.~\ref{sect:stability}).
Thermal pulses, if they occur, are considerably weakened  with rotation. 

As mentioned previously, and as discussed by Cassisi et al. (\cite{Cassisi98}), 
Kato \& Hachisu (\cite{Kato99}) and Langer et al. (\cite{Langer02}), 
a high mass accumulation efficiency
on a white dwarf by hydrogen accretion has been seriously questioned
due to the unstable helium shell burning which is found 
when helium is accreted at
a rate  which  gives steady hydrogen shell burning.
Note that the accretion rate of \hemdotc{} \msyr{} corresponds roughly 
to the maximum rate for the steady hydrogen burning 
in a 1.0 \Msun{}  white dwarf (Nomoto~\cite{Nomoto82a}). 
Our results indicate that even if hydrogen is
accreted and burned steadily, the subsequent helium shell burning
may be unstable, with strong helium shell flashes driving the stellar envelope
to the Eddington limit, unless effects of rotation are considered.
Kato \& Hachisu (\cite{Kato99}) suggested that 
if the new OPAL opacities are adopted, an optically thick wind
can remove mass from the system without the white dwarf filling its Roche lobe. 
They also suggested that the effective mass accumulation rate can remain
high enough for the white dwarf to grow to the Chandrasekhar limit.
However, their results are based on a very massive white dwarf model of 1.3 \Msun{}. 
Cassisi et al. (\cite{Cassisi98}), on the other hand, showed
that hydrogen accreting white dwarfs with  initial masses of 0.56 \Msun{} and 0.8 \Msun{}  
fill their Roche lobe 
once helium flashes occur,
leaving doubts about the efficient mass increase of the white dwarf 
from a relatively low mass 
by hydrogen accretion.

Our results imply that rotation may help white dwarfs
to grow in mass by stabilizing helium shell burning.
In the canonical single degenerate 
scenario, however,
a main sequence star or a red giant is supposed to
transfer hydrogen rich matter onto the white dwarf
(Li \& van den Heuvel~\cite{Li97}; Hachisu et al.~\cite{Hachisu99}; 
Langer et al.~\cite{Langer00}).
We still need to investigate the case of hydrogen accretion, 
in order to answer the question whether rotation may be a potential solution
in explaining the observed SNe~Ia rate from the considered binary systems.
Since the stability conditions for hydrogen shell burning
are not qualitatively different from that for helium shell burning
(Yoon, Langer \& van der Sluys~\cite{Yoon04d}), 
we expect that the effects of rotation on the hydrogen shell source 
are likely to give a similar conclusion 
as for the helium shell burning. 

On the other hand, hydrogen accretion with a rate of  
\hemdotd{} \msyr{} is supposed to expand a 1.0~\Msun{}
white dwarf into a red giant phase (Nomoto~\cite{Nomoto82a}; Nomoto \& Kondo~\cite{Nomoto91};
Cassisi et al.~\cite{Cassisi98}). 
In a close binary system, the hydrogen envelope expanded to the red giant phase
will fill the Roche lobe, causing significant mass loss or the merging of both stars,
unless the mass accretion rate is not reduced by an optically
thick stellar wind, as suggested by Hachisu et al (\cite{Hachisu96},~\cite{Hachisu99}). 
However, steady helium accretion with such a high accretion rate can be realized, for example, 
in a binary system which consists of a white dwarf and a helium giant star 
(``helium Algols'', Iben \& Tutukov~\cite{Iben94}), in which
the mass transfer rate can amount to
$10^{-6} - 10^{-5}$ \msyr{}, as shown by Yoon \& Langer (\cite{Yoon03}). 
Our results indicate clearly that the mass accumulation 
efficiency can be significantly enhanced by rotation for helium giant star + 
white dwarf binary systems.
Although the calculated production rate of SNe Ia via this route is currently 
rather uncertain (Iben \& Tutukov~\cite{Iben94}; Branch et al.~\cite{Branch98}), 
consideration of rotation may increase it strongly.

In the present study, we did not consider stellar winds from white dwarfs,
on which rotation might cause significant effects. When the stellar surface approaches
the critical velocity, the mass loss due to radiation-driven winds could
increase dramatically (Friend \& Abbott~\cite{Friend86}; Langer~\cite{Langer97}). 
Since stellar winds may also carry a large amount of angular momentum 
(Langer~\cite{Langer98}), 
this would involve a complicated history 
of angular momentum gain or loss in accreting white dwarfs:
accretion will spin up white dwarfs, while stellar winds will brake them.
It is also possible that the physical conditions in the shell source
in a rapidly rotating white dwarf
vary with latitude, which may lead to interesting
consequences such as non-spherical stellar winds 
(cf. Owocki \& Gayely~\cite{Owocki97}; Maeder~\cite{Maeder99}). 
We will investigate these effects systematically in near future.

Finally, we note that an increase in the uncertain efficiency of 
rotationally induced chemical mixing results in more stable 
shell burning (Fig.~\ref{fig:lhe1} and~\ref{fig:lhe2}), than
the case where the calibration by Heger et al. (\cite{Heger00a}) and
Heger \& Langer (\cite{Heger00b}) is adopted. 
Their calibration gives
a similar chemical mixing efficiency as in
Pinsonneault et al. (\cite{Pinsonneault89}) who made a calibration to reproduce 
the solar surface \Li{7} abundance, 
as well as to the theoretical estimate by Chaboyer \& Zahn (\cite{Chaboyer92}). 
The physical conditions in the helium shell source in white dwarfs 
do not qualitatively differ from those in main sequence stars. 
However, for accreting white dwarfs, 
another mixing mechanism which was not considered in the present study
may be important: magnetic instability induced by 
differential rotation in radiative layers (Spruit~\cite{Spruit02}).
Heger et al. (\cite{Heger03}) and Maeder \& Meynet (\cite{Maeder03b}) show that 
the turbulent diffusion can become stronger by several 
orders of magnitude
due to the magnetic instability, compared to 
the case where only rotational effects are considered.
The differential rotation induced magnetic instability 
can be even more important in accreting white dwarfs,
in which a strong shear motion is retained 
due to the continuous angular momentum 
transport from the accreted matter, as shown in Sect.~\ref{sect:spin}.
The efficiency of the chemical mixing may increase accordingly,
which may stabilize the shell source even further than in the present study.
This possibility is currently under investigation.

\begin{acknowledgements}
We would like to thank the anonymous referee for useful comments
and careful reading of the manuscript, which led to significant
improvement of the text.
We are grateful to Peter H\"oflich for enlightening discussions.
This research has been supported in part by the Netherlands Organization for
Scientific Research (NWO).
\end{acknowledgements}

\end{document}

%% file: ndef.tex
 \def\h1{\hangindent=1.0truecm \hangafter=0}                                    
 \def\simle{\mathrel{\hbox{\rlap{\hbox{\lower4pt\hbox{$\sim$}}}\hbox{$<$}}}}    
 \def\simgr{\mathrel{\hbox{\rlap{\hbox{\lower4pt\hbox{$\sim$}}}\hbox{$>$}}}}    

 \def\hb{\hfill\break}
 \def\h2{\hb\noindent \hangindent=0.5cm \hangafter=1}

 \def\utw{\smash{\rlap{\lower5pt\hbox{$\sim$}}}}
 \def\udtw{\smash{\rlap{\lower6pt\hbox{$\approx$}}}}

 \def\apjl{Ap.~J.~ Letters~}

%% file: shell.bbl
\begin{thebibliography}{}
\bibitem[1995]{Branch95} Branch, D., Livio, M., Yungelson, L.R., Boffi, F.R., \& Baron, E., 1995,  PASP, 107, 1019
\bibitem[1998]{Branch98} Branch, D., 1998, ARA\&A, 36, 17
\bibitem[1998]{Cassisi98} Cassisi, S., Iben, I., \& Tornamb\'e, A., 1998, ApJ, 496, 376 
\bibitem[1985]{Caughlan85} Caughlan, G.R., Fowler, W.A., Marris, M.J., \& Zimmerman, B.A., 1985, Atomic Dat. Nuc. Dat. Tables., 32, 197
\bibitem[1992]{Chaboyer92} Chaboyer, B., \& Zahn, J.-P., 1992, A\&A, 253, 173
\bibitem[1970]{Chandrasekhar70} Chandrasekhar, S. 1970, Phys. Rev. L, 24, 611
\bibitem[1968]{Clayton68} Clayton, D.D. 1968, Principles of Stellar Evolution and Nucleosynthesis (New York: MacGraw-Hill)
\bibitem[2001]{Dominguez01} Dom\'inguez, I., H\"oflich, P., \& Straniero, O., 2001, ApJ, 557, 279
\bibitem[1975]{Durisen75b} Durisen, R.H., 1975, ApJ, 199, 179
\bibitem[1977]{Durisen77} Durisen, R.H., 1977, ApJ, 213, 145
\bibitem[1981]{Durisen81} Durisen, R.H., \& Imamura, J.N., 1981, ApJ, 243, 612
\bibitem[1976]{Endal76} Endal, A.S., \& Sofia, S., 1976, ApJ, 210, 184
\bibitem[1993]{Fliegner93} Fliegner, J., 1993, Diplomarbeit, University of G\"ottingen
\bibitem[1978]{Friedman78} Friedman, J.L., \& Schutz, B., 1978, ApJ, 222, 281
\bibitem[1986]{Friend86} Friend, D.S., \& Abbott, D.C., 1986, ApJ, 311, 701
\bibitem[1988]{Fujimoto88} Fujimoto, M.Y., 1988, A\&A, 198, 163 
\bibitem[1997]{Fujimoto97} Fujimoto, M.Y., \& Iben I., 1997, in 
      Advances in Stellar Evolution, Cambridge University Press.
\bibitem[1979]{Fujimoto79} Fujimoto, M.Y., \& Sugimoto, D., 1979, PASJ, 31, 1 
\bibitem[2000]{Greiner00} Greiner, J., 2000, New Astron., 5, 137
\bibitem[1986]{Hachisu86} Hachisu, I., 1986, ApJS, 62, 461
\bibitem[2001]{Hachisu01} Hachisu, I., \& Kato, M., 2001, ApJ, 558, 323
\bibitem[1996]{Hachisu96} Hachisu, I., Kato, M., \& Nomoto, K., 1996, ApJ, 470, L97
\bibitem[1999]{Hachisu99} Hachisu, I., Kato, M., \& Nomoto, K., 1999, ApJ, 522, 487
\bibitem[1996]{Hamuy96a} Hamuy, M., Phillips, M.M., \& Suntzeff, N.B., et al., 1996, ApJ, 519, 314
\bibitem[1997]{Heber97} Heber, U., Napiwotzki, R., \& Reid, I.N., 1997, A\&A, 323, 819
\bibitem[2000]{Heger00b} Heger, A., \& Langer, N, 2000, ApJ, 544, 1016
\bibitem[2000]{Heger00a} Heger, A., Langer, N., \& Woosley, S.E., 2000, ApJ, 528, 368
\bibitem[2003]{Heger03} Heger, A.,  Woosley, S., Langer, N., \& Spruit, H.C., 2003, in: Stellar Rotation, proc. IAU-Symp. 215, (San Francisco: ASP), A. Maeder, P. Eenens, eds., in press
\bibitem[2000]{Hillebrandt00} Hillebrandt, W., \& Niemeyer, J.C., 2000, ARA\&A, 38, 191
\bibitem[1996]{Hoeflich96b} H\"oflich, P., Khokhlov, A., \& Wheeler, J.C., et al.,  1996, ApJ, 472, 81
\bibitem[2000]{Hoeflich00} H\"oflich, P., Nomoto, K., Umeda, H., \& Wheeler, J.C., 2000, ApJ, 528, 590 
\bibitem[2001]{Howell01} Howell, D.A., H\"oflich, P., Wang, L., \& Wheeler, J.C., 2001, ApJ, 556, 302
\bibitem[1995]{Hujeirat95} Hujeirat, A., 1995, A\&A, 295, 268
\bibitem[1983]{Iben83a} Iben, I.Jr., \& Renzini, A., 1983, ARA\&A, 27, 271
\bibitem[1991]{Iben91} Iben, I.Jr., \& Tutukov, A.V., 1991, ApJ, 370, 615
\bibitem[1994]{Iben94} Iben, I.Jr., \& Tutukov, A.V., 1994, ApJ, 431, 264
\bibitem[1996]{Iglesias96} Iglesias, C.A., \& Rogers, F.J., 1996, ApJ, 464, 943
\bibitem[1995]{Imamura95} Imamura, J.N., Toman, J., Durisen, R.H., Pickett, B., \& Yang, S., 1995, ApJ, 444, 363
\bibitem[1997]{Kahabka97} Kahabka, P., \& van den Heuvel, E.P.J., 1997, ARA\&A, 35, 69
\bibitem[2003]{Karl03} Karl, C.A., Napiwotzki, R., \& Nelemans, G. et al., 2003, A\&A, 410, 663
\bibitem[2003]{Kasen03} Kasen, D., Nugent, P., \& Wang, L., et al., 2003, ApJ, 593, 788
\bibitem[1999]{Kato99} Kato, M., \& Hachisu, I., 1999, ApJ, 513, L41
\bibitem[2003]{Kawaler03} Kawaler, S.D., 2003, in: Stellar Rotation, proc. IAU Symp. 215, A. Maeder, \& P. Eenens, eds.
\bibitem[1974]{Kippenhahn74} Kippenhahn, R., \& M\"ollenhoff, C., 1974, Ap.\&SS, 31, 117
\bibitem[1970]{Kippenhahn70} Kippenhahn, R., \& Thomas, H.-C., 1970, In: Stellar Rotation, ed. A. Slettebak, IAU Coll 4, Reidel, Dortrecht
\bibitem[1978]{Kippenhahn78} Kippenhahn, R., \& Thomas, H.-C., 1978, A\&A, 63, 265
\bibitem[1998]{Koester98} Koester, D., Dreizler, S., Weidemann, V., \& Allard, N.F., 1998, A\&A, 338, 617
\bibitem[2001]{Koester01} Koester, D., Napiwotzki, R., \& Christlieb, N., et al., 2001, A\&A, 378, 556
\bibitem[1997]{Langer97} Langer, N., 1997, A\&A, in Luminous Blue Variables: Massive Stars in Transition, Nota, A, \& Lamers,  H.J.G.L.M., eds, ASPC, 120, p. 83
\bibitem[1998]{Langer98} Langer, N., 1998, A\&A, 329,551
\bibitem[1999]{Langer99} Langer, N., Heger A., Wellstein, S., \& Herwig, F., 1999, A\&A, 346, L37
\bibitem[2000]{Langer00} Langer, N., Deutschmann, A., Wellstein, S., \& H\"oflich, P., 2000, A\&A, 362, 1046
\bibitem[2002]{Langer02} Langer, N., Yoon, S.-C., Wellstein, S., \& Scheithauer, S. 2002, 
In: The Physics of Cataclysmic Variables and Related Objets, ASP Conference Proceedings, vol. 261
\bibitem[2001]{Leibundgut01} Leibundgut, B, 2001, ARA\&A, 39, 67
\bibitem[1997]{Li97} Li X.-D., \& van den Heuvel, E.P.J., 1997, A\&A, 322, L9
\bibitem[1991]{Limongi91} Limongi, M., \& Tornamb\'e, A., 1991, ApJ, 371, 317
\bibitem[1999]{Lindblom99} Lindblom, L, 1999, Phys. Rev. D, 60, 4007  
\bibitem[2001]{Livio01} Livio, M., 2001, In: Cosmic evolution, Vangioni, E.,  Ferlet, R, \&  Lemoine, M., eds New Jersey: World Sceintific
\bibitem[1998]{Livio98} Livio, M., \& Pringle, J.E., 1998, ApJ, 505, 339
\bibitem[1987]{Livio87} Livio, M.,  \& Truran, J., 1987, ApJ, 318, 316
\bibitem[1995]{Livne95} Livne, E.,  \& Arnett, D., 1995, ApJ, 452, 62
\bibitem[1983]{MacDonald83} MacDonald, J.,1983, ApJ, 273,289
\bibitem[1999]{Maeder99} Maeder, A., 1999, A\&A,  347, 185
\bibitem[2003]{Maeder03a} Maeder, A., 2003, A\&A,  399, 263
\bibitem[2003]{Maeder03b} Maeder, A,  \& Meynet, G., 2003, A\&A, 411, 543
\bibitem[2000]{Maxted00} Maxted, P.F.L., March, T.R., \& North, R.C., 2000, MNRAS, 317, L41
\bibitem[1997]{Meynet97} Meynet, G., \& Maeder, A., 1997, A\&A, 321, 465
\bibitem[1982a]{Nomoto82a} Nomoto, K., 1982a, ApJ, 253, 798
\bibitem[1982b]{Nomoto82b} Nomoto, K., 1982b, ApJ, 257, 780
\bibitem[1997]{Nomoto97b} Nomoto, K., Iwamoto, K., \& Kishimoto, N., 1997, Science, 276, 1378
\bibitem[1991]{Nomoto91} Nomoto, K., \& Kondo, Y., 1991, ApJ, 367, L19
\bibitem[1997]{Nugent97} Nugent, P., Baron, E., Branch, D., Fisher, A., \& Hauschildt, P.H., 1997, ApJ, 485, 812
\bibitem[1968]{Ostriker68a} Ostriker, J.P., \& Bodenheimer, P., 1968, ApJ, 151, 1089
\bibitem[1968]{Ostriker68b} Ostriker, J.P., \& Mark, J.W.-K., 1968, ApJ 151, 1075
\bibitem[1997]{Owocki97} Owocki, S.P., \& Gayley, K.G., 1997, In: Luminous Blue Variables:Massive stars in Transition, Nota, A., \& Lamers, H.J.G.L.M., eds.,  ASP Conf. Ser. Vol. 120
\bibitem[1991]{Paczynski91} Paczy\'nski, B., 1991, ApJ, 370, 597
\bibitem[1999]{Perlmutter99a} Perlmutter, S., Aldering, G., \& Goldhaber, G., et al., 1999, ApJ, 517, 565
\bibitem[1989]{Pinsonneault89} Pinsonneault, M.H., Kawaler, S.D., Sofia, S., \& Demarque, P., 1989, ApJ, 338, 424
\bibitem[2001]{Pinto01} Pinto, P.A., Eastman, R.G., \& Rogers, T., 2001, ApJ, 551, 231
\bibitem[1991]{Popham91} Popham, R., \& Narayan, R., 1991, ApJ, 370, 604
\bibitem[2000]{Riess00} Riess, A.G., Filippenko, A.V., \& Liu, M.C., et al., 2000, ApJ, 536, 62
\bibitem[1985]{Ritter85} Ritter, H., 1985, A\&A, 148, 207
\bibitem[1985]{Saio85} Saio, H., \& Nomoto, K., 1985, A\&A, 150, L21
\bibitem[1998]{Saio98} Saio, H., \& Nomoto, K., 1998, ApJ, 500, 388
\bibitem[1999]{Sion99} Sion, E.M., 1999, PASP, 111,532
\bibitem[1987]{Sparks87} Sparks, W.M., \& Kutter, G.S., 1987, ApJ, 321, 394
\bibitem[1965]{Schwarzshild65} Schwarzshild, M., \& H\"arm, R., 1965, ApJ, 142, 855
\bibitem[2003]{Starrfield03} Starrfield, 2003, in: Stellar Rotation, proc. IAU Symp. 215, A. Maeder, P. Eenens, eds.
\bibitem[2002]{Spruit02} Spruit, H.C., 2002, A\&A, 381, 923
\bibitem[2001]{Thoroughgood01} Thoroughgood, T.D., Dhillon, V.S., Littlefair, S.P., Marsh, T.R., \& Smith, D.A.,2001, 
MNRAS, 327, 1323
\bibitem[2001]{Tout01} Tout, C.A., Reg\"os, E., Wickramasinghe, D., Hurley, J., \& Pols, O.R., 2001, In: Evolution
of Binary and Multiple Star Systems, ASP Conf. Ser., Vol 229, Ph. Podsialdlowski et al., eds.
\bibitem[1999]{Umeda99} Umeda, H., Nomoto, K., Kobayashi, C., Hachisu, I., \& Kato M., 1999, ApJ, 522, L43
\bibitem[2003]{Wang03} Wang, L., Baade, D., \& H\"oflich, P., et al., 2003, ApJ, 591, 1110
\bibitem[1993]{Weaver93} Weaver, T.A., \& Woosley, S.E., 1993, Phys. Rep., 227, 65 
\bibitem[1966]{Weigert66} Weigert, A., 1966, Z. Astrophys., 64, 395
\bibitem[1996]{Wheeler96} Wheeler,  C.J., 1996, In: Evolutionary Processes in Binary Stars, Dordrecht, Kluwer 
\bibitem[1986]{Woosley86} Woosley, S.E., \& Weaver, T.A., 1986, ARA\&A, 24, 205
\bibitem[1994]{Woosley94} Woosley, S.E., \& Weaver, T.A., 1994, ApJ, 423, 371
\bibitem[2002]{Yoon02} Yoon, S.-C., \& Langer, N., 2002
In: The Physics of Cataclysmic Variables and Related Objets, ASP Conference Proceedings, vol. 261,  
B.T. Gaensicke et al., eds.
\bibitem[2003]{Yoon03} Yoon, S.-C., \& Langer, N., 2003, A\&A, 412, L53
\bibitem[2004a]{Yoon04a} Yoon, S.-C., \& Langer, N., 2004a, A\&A, 419, 623 
\bibitem[2004b]{Yoon04b} Yoon, S.-C., \& Langer, N., 2004b, A\&A, 419, 645
\bibitem[2004]{Yoon04d} Yoon, S.-C., Langer, N., \& van der Sluys, M., 2004, A\&A, in press
\bibitem[1975]{Zahn75} Zahn, J.-P., 1975, Mem. Soc. Roy. Li\`ege, 8, 31
\bibitem[1992]{Zahn92} Zahn, J.-P., 1992, A\&A, 265, 115

\end{thebibliography}
